\documentclass[10pt,twocolumn,letterpaper]{article}

\usepackage{multirow}
\usepackage{graphicx} 
\usepackage{float} 
\usepackage{array}
\usepackage{svg}
\usepackage{subcaption}
\usepackage{colortbl} 
\usepackage{amssymb}
\usepackage{pifont}
\usepackage{fontawesome}
\usepackage{siunitx}

\definecolor{lightblue}{RGB}{197, 234, 228} 
\newcommand{\cmark}{\textcolor{green}{\ding{51}}} 
\newcommand{\xmark}{\textcolor{red}{\ding{55}}}   

\usepackage[pagenumbers]{cvpr} 

\definecolor{cvprblue}{rgb}{0.21,0.49,0.74}
\usepackage[pagebackref,breaklinks,colorlinks,allcolors=cvprblue]{hyperref}

\def\paperID{11939} 
\def\confName{CVPR}
\def\confYear{2025}

\title{Do Audio-Visual Segmentation Models Truly Segment Sounding Objects?}

\author{
Jia Li$^{1}$ \hspace{1cm} Wenjie Zhao $^{1}$ \hspace{1cm} Ziru Huang$^{2}$ \hspace{1cm} Yunhui Guo$^{1}$ \hspace{1cm} Yapeng Tian$^{1}$ \\
$^{1}$ The University of Texas at Dallas \hspace{1cm}
$^{2}$ Tsinghua University\\
$^{1}${\tt\small  \{Jia.Li, Wenjie.Zhao,  Yunhui.Guo, Yapeng Tian\}@utdallas.edu} $^{2}$ {\tt\small huangzr21@mails.tsinghua.edu.cn}
}

\begin{document}
\maketitle

\begin{abstract}
\vspace{-5pt}

Unlike traditional visual segmentation, audio-visual segmentation (AVS) requires the model not only to identify and segment objects but also to determine whether they are sound sources.
Recent AVS approaches, leveraging transformer architectures and powerful foundation models like SAM, have achieved impressive performance on standard benchmarks. Yet, an important question remains: Do these models genuinely integrate audio-visual cues to segment sounding objects?
In this paper, we systematically investigate this issue in the context of robust AVS. Our study reveals a fundamental bias in current methods: they tend to generate segmentation masks based predominantly on visual salience, irrespective of the audio context. This bias results in unreliable predictions when sounds are absent or irrelevant.
To address this challenge, we introduce AVSBench-Robust, a comprehensive benchmark incorporating diverse negative audio scenarios including silence, ambient noise, and off-screen sounds. We also propose a simple yet effective approach combining balanced training with negative samples and classifier-guided similarity learning.
Our extensive experiments show that state-of-the-art AVS methods consistently fail under negative audio conditions, demonstrating the prevalence of visual bias. In contrast, our approach achieves remarkable improvements in both standard metrics and robustness measures, maintaining near-perfect false positive rates while preserving high-quality segmentation performance. 

\vspace{-3mm}

\end{abstract}

\section{Introduction}
\label{sec:intro}

\begin{figure}[h]
    \centering
    \includegraphics[width=0.48\textwidth]{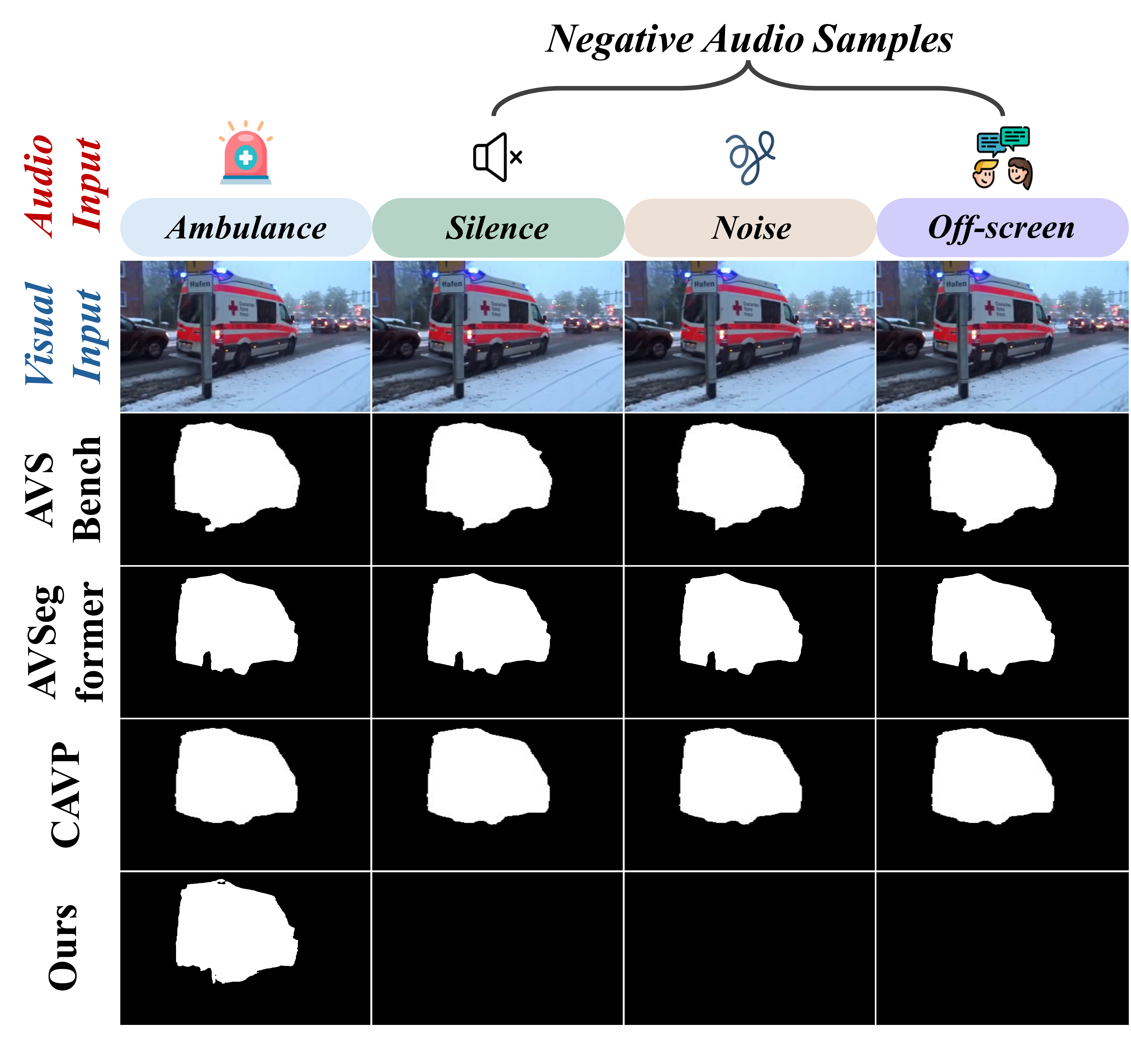}
    \vspace{-7mm}
    \caption{\textbf{Performance in Different Audio Scenarios.} The top row shows an ambulance image under different audio conditions: \textit{Ambulance sound} (positive), \textit{Silence}, \textit{Noise}, and \textit{Offscreen sounds} (negative). Each subsequent row displays the segmentation output various SOTA AVS models~\cite{zhou2022audio, gao2024avsegformer, chen2024cavp} and our model under each audio condition. In negative scenarios, existing models segment the ambulance due to ``visual prior" bias, mistakenly associating it with unrelated audio. In contrast, our model accurately segments only in the presence of relevant audio, demonstrating improved alignment between audio cues and visual segmentation.
    }
    \vspace{-7mm}

    \label{fig:fig1}
\end{figure}

Audio-Visual Segmentation (AVS) aims to identify and segment sounding objects within visual scenes~\cite{zhou2022audio,gao2024avsegformer}. This essential multimodal task mirrors a fundamental aspect of human perception: the integration of auditory and visual stimuli to focus attention on relevant sources~\cite{zhou2022audio, small2005odor, chen2020vggsound}. For instance, when hearing a baby cry, people naturally locate the sound’s visual source.  Simulating this ability in machines could open up valuable cross-modal applications, such as improved multimedia analysis~\cite{arandjelovic2017look, arandjelovic2018objects, hu2022mix, mo2024multi}, enhanced human-computer interaction~\cite{yang2024analyzing, wang2024audiobench, johansen2022characterising, lv2022deep, fu2021design}, and autonomous systems capable of interpreting sound-emitting objects in complex environments~\cite{schmidt2020acoustic, dutt2020self, topcu2020assured}. 

Recent years have witnessed remarkable progress in AVS.  State-of-the-art (SOTA) methods leverage multimodal information, utilizing encoder-decoder structures with audio-visual interaction~\cite{zhou2022audio}, multimodal transformer architectures~\cite{gao2024avsegformer,li2024qdformer,liu2023AuTR}, audio query-guided designs~\cite{liu2023AuTR, sun2024biasinAVS}, and strong vision foundation models~\cite{mo2023av,liu2024annofree, wang2024GAVS, sun2024biasinAVS} like SAM~\cite{kirillov2023sam}and Mask2Former~\cite{cheng2022mask2former}. 
These innovations have driven impressive performance on standard benchmark datasets: AVSBench-S4 and AVSBench-MS3 ~\cite{zhou2022audio}.

However, a critical question arises: \textit{are these models truly performing audio-visual segmentation, or simply conducting visual segmentation with minimal audio integration?}  AVS, by definition, introduces a crucial constraint: only objects acting as sound sources should be segmented. For example, an AVS model should not segment a visually salient yet silent dog.  Current AVS models, primarily trained and evaluated on ``positive'' cases where visual objects correspond to audio cues, often neglect scenarios with unrelated sounds, such as silence or off-screen sources.


To systematically investigate whether AVS models truly integrate audio-visual information, we introduce AVS-Robust, a comprehensive benchmark comprising 4,932 single-source and 424 multi-source videos across 20 diverse object classes from AVSBench~\cite{zhou2022audio}. We incorporate four different audio conditions for each video: original audio, silence, ambient noise, and off-screen sounds. Each condition represents 25\% of the evaluation scenarios. Our study reveals a concerning bias in existing SOTA methods: they tend to generate segmentation masks based primarily on visual salience, irrespective of the audio context. For instance, these models may segment an ambulance even in the presence of silence or unrelated ambient sounds, indicating an over-reliance on visual cues rather than genuine audio-visual integration (Fig.~\ref{fig:fig1}).



Building upon these insights, we explore solutions to address this visual bias. While incorporating negative audio-visual pairs during training seems intuitive, this approach alone presents a challenge: without explicit guidance for audio-visual integration, models struggle to determine whether to segment objects based solely on visual information. To overcome this, we propose a debiasing approach with two key components:
\textit{(1) Balanced Training with Negative Samples:} Incorporating both positive and negative audio-visual pairs during training to expose models to a wider range of audio-visual relationships. \textit{(2) Classifier-Guided Similarity Learning:} Utilizing a classifier to guide the model in learning effective audio-visual feature representations and promoting similarity between corresponding audio and visual features.

Extensive experiments using our new benchmark yield several crucial findings. Recent SOTA methods, including SAMA-AVS \cite{liu2024annofree}, Stepping-Stones \cite{ma2024stepping}, and CAVP \cite{chen2024cavp}, consistently fail under negative audio conditions, exhibiting high False Positive Rates (FPR). When evaluated with our comprehensive metrics—such as G-mIoU, G-F, and G-FPR, as discussed in Sec.~\ref{sec:problem_benchmark}—these models show significant performance degradation compared to their reported results on standard benchmarks. In contrast, our approach achieves superior performance across all robustness metrics while maintaining competitive segmentation quality on positive audio inputs in both single- and multi-source scenarios.

Our main contributions are summarized as follows:
\begin{itemize}    
    \item  We conduct a systematic study on audio robustness in AVS and introduce AVSBench-Robust along with our new robustness evaluation protocols. This benchmark rigorously evaluates AVS models under both standard conditions and challenging negative scenarios, assessing their ability to effectively integrate audio-visual information.
    \item We propose a training strategy for robust AVS by incorporating diverse negative audio scenarios and employing classifier-guided similarity learning, which enhances model robustness and preserves segmentation quality.
    \item Extensive experiments demonstrate that our approach substantially outperforms current SOTA methods in terms of our robustness metrics while achieving competitive performance on standard AVS benchmarks. 
\end{itemize}

\section{Related work}
\label{sec:related_new}
\vspace{-4pt}

\textbf{Sound Source Localization.}
This task is closely related to AVS, focusing on localizing sound sources within visual scenes \cite{arandjelovic2018objects, senocak2018learning,mo2022closer, chen2021localizing,mahmud2024t}. This task advances cross-modal understanding through various technical approaches, from basic feature fusion strategies to sophisticated attention mechanisms \cite{senocak2018learning, mo2022closer, hu2020discriminative, qian2020multiple}.
Recent sound source localization approaches have significantly improved sound source discrimination through multiple innovations: contrastive learning with hard-mining strategies enhances complex region distinction~\cite{chen2021localizing, hu2020discriminative, mo2022closer}, while class-aware approaches and dual-phase feature alignment enable robust multi-source localization without explicit pairwise annotations \cite{hu2020discriminative, qian2020multiple, chen2021localizing}.
However, the predicted sounding object heatmaps lack the fine-grained precision offered by AVS's pixel-level segmentation capabilities.

\vspace{2mm}
\noindent
\textbf{Audio-Visual Segmentation.}
AVS task focuses on identifying and segmenting sound-producing objects through pixel-level mask prediction. 
The field has progressed significantly since its inception~\cite{zhou2022audio,gao2024avsegformer, li2023catr, chen2024cavp, liu2024annofree, ma2024stepping, wang2024GAVS, guo2024open, sun2024biasinAVS}. Most approaches follow an encoder-decoder design, with early works focusing on effective fusion strategies for audio-visual information~\cite{zhou2022audio}. Subsequent developments explored more sophisticated architectures, incorporating multimodal transformers and audio query guided mechanisms~\cite{gao2024avsegformer, liu2023AuTR, sun2024biasinAVS, wang2024ref, li2023catr, ma2024stepping, yang2024combo} to enhance cross-modal understanding. Recently, methods leveraging vision foundation models~\cite{liu2024annofree, wang2024GAVS, sun2024biasinAVS} and LLMs~\cite{wang2024can, he2024mlmseg}  have demonstrated improved segmentation capabilities. 


Despite these developments, we observe that AVS models commonly suffer from ``visual prior'' bias, where models generate predictions primarily based on visual salience regardless of audio context~\cite{sun2024biasinAVS, chen2024cavp, li2024qdformer}. While recent efforts address this through contrastive learning~\cite{chen2024cavp} and semantic enhancement~\cite{sun2024biasinAVS, li2024qdformer}, they still lead to disregard audio features (see Fig.~\ref{fig:fig1}) or face limited evaluation scope. 

In contrast to complex architectural solutions, we adopt a straightforward yet effective approach to enhancing AVS model robustness. We systematically analyze audio robustness in AVS and introduce AVSBench-Robust, a benchmark designed to evaluate models under both original audio conditions and challenging negative audio scenarios. 
A concurrent study \cite{juanola2024critical} identified visual bias in SSL models, introducing the same negative audio scenarios along with evaluation metrics, but primarily focuses on assessing model performance.
In contrast, our work not only evaluates robustness but also presents a targeted training strategy to strengthen AVS models against misleading audio-visual cues.

\vspace{2mm}
\noindent
\textbf{Imbalanced Multimodal Learning.}
Recent studies in audio-visual learning highlight significant challenges in balancing different modalities during training, where dominant modalities often overshadow others in the learning process \cite{wang2020mmlhard,tian2020unified,wei2024mmpareto, peng2022balanced}. Various solutions such as modality-specific optimization, gradient modulation, and Pareto optimization~\cite{wang2020mmlhard, wei2024mmpareto, peng2022balanced} have been proposed to address this imbalance, aiming to preserve the contribution of each modality. However, these approaches primarily target tasks where separate losses are applied for joint modalities and each individual modality ~\cite{wang2020mmlhard, wei2024mmpareto}. In contrast, the AVS task typically involves applying a single loss after fusing audio and visual features, which requires us to design a new strategy for effective modality balancing in AVS.

\section{Problem and Benchmark}
\label{sec:problem_benchmark}


In this section, we first present the formulation of the AVS task and its associated challenges in Sec.~\ref{subsec:task}. In Sec.~\ref{subsec:benchmark}, we introduce our new benchmark for robust AVS. Finally, we present our evaluation protocols in Sec.~\ref{subsec:eval}.

\subsection{Task and Challenges}
\label{subsec:task}
Given \( T \) non-overlapping video and audio clips \( \{V^t, A^t\}_{t=1}^{T} \), the goal of the AVS task is to predict a segmentation mask \( \mathcal{M_{\text{pred}}}^t \in \mathbb{R}^{H \times W} \) that labels sounding pixels in each video frame of the clips, where \( H \) and \( W \) denote the frame dimensions, and the mask is binary. Following previous studies~\cite{zhou2022audio,gao2024avsegformer, li2023catr}, we extract a single video frame at the end of each second and set \( T = 5 \) in practice, so each clip contains only one extracted frame. 

Unlike purely visual segmentation, AVS inherently addresses two subtasks simultaneously: segmenting visual objects and determining whether they are sound sources. Therefore, predictions should satisfy two key requirements: (1) when objects are producing sound, the model should generate accurate segmentation masks for those objects, and (2) when no audio-visual correspondence exists, the model should produce empty masks to avoid false predictions.



While current AVS approaches have shown promising segmentation results on standard benchmark protocols, they face a fundamental limitation: evaluations have primarily focused on positive cases where audio and visual signals fully align, with salient objects in video frames typically being sound sources in the datasets. This setup allows AVS models to potentially rely solely on visual information to achieve high performance, bypassing true multimodal integration.
This focus on positive cases overlooks the equally important ability to suppress predictions when no valid audio-visual correspondence exists. To comprehensively assess the multimodal learning capabilities of AVS models, a more robust benchmark is needed.




\begin{table}[t]
\centering
\footnotesize
\vspace{-5mm}
\renewcommand{\arraystretch}{1.1}  
\setlength{\tabcolsep}{10pt}  

\begin{tabular}{>{\raggedright\arraybackslash}m{0.15\linewidth}|>{\raggedright\arraybackslash}m{0.65\linewidth}}
\hline
\textbf{Category} & \textbf{Representative  Examples} \\
\hline
\faMusic \textbf{ Music} & Guitar, Violin, Piano, Tabla, Marimba, Ukulele, Playing Acoustic Guitar, Playing Glockenspiel, Playing Violin, Playing Ukulele \\
\hline
\faChild \textbf{ Human Voice} & Male Speech, Female Speech, Male Singing, Female Singing, Baby Crying, Baby Laughter \\
\hline
\faPaw \textbf{ Animals} & Dog Barking, Lion Roaring, Cat Meowing, Bird Chirping, Wolf Howling, Horse Neighing, Coyote Howling, Mynah Bird Singing \\
\hline
\faCar \textbf{ Devices, Machines} & Helicopter, Ambulance Siren, Car Horn, Lawn Mower, Chainsaw, Bus Engine, Typing on Computer Keyboard, Cap Gun Shooting, Emergency Car, Driving Buses, Race Car \\
\hline
\end{tabular}
\caption{Semantic Categories and Examples in AVSBench-Robust: To ensure clear evaluation of cross-modal understanding, we organize sounds into distinct semantic categories, which particularly crucial for the off-screen audio condition.}
\label{tab:sound-categories}
\vspace{-5mm} 

\end{table}
\subsection{AVSBench-Robust}  
\label{subsec:benchmark}

To address the limitations of current evaluation frameworks and facilitate the development of more robust AVS models, we introduce {AVSBench-Robust}. Building upon AVSBench~\cite{zhou2022audio}, this benchmark includes two evaluation scenarios: (1) the single-source subset (S4), containing 4,932 videos (3,452 for training, 740 for validation, and 740 for testing), and (2) the multi-source subset (MS3), comprising 424 videos (296 for training, 64 for validation, and 64 for testing). For each video from AVSBench~\cite{zhou2022audio}, we create three additional negative audio conditions alongside the original positive audio, effectively quadrupling the number of audio-visual pairs for a comprehensive evaluation.

The benchmark spans 20 diverse sound-producing object classes across four major categories: machine (32.2\%), music (32.1\%), animal (23.2\%), and human (12.5\%). To evaluate model robustness, each video is paired with four types of audio conditions:

\noindent
\underline{\textit{Positive Pair:}} Original audio of the video from AVSBench, where the audio accurately reflects the visible objects.

\noindent
\underline{\textit{Silence Scenario:}} Test cases without audio, where objects are visually present but silent in the video.

\noindent
\underline{\textit{Noise Condition:}} Background audio noise, testing the model's ability to differentiate between meaningful and irrelevant audio signals.

\noindent
\underline{\textit{Off-screen Audio:}} Semantically unrelated sounds from different categories, as outlined in Table~\ref{tab:sound-categories}, testing the model’s ability to maintain accurate audio-visual correspondence. For example, pairing animal visuals with device sounds requires models to learn true cross-modal relationships rather than relying solely on visual cues.

By incorporating unpaired audio clips, we create negative audio-visual pairs that enable the study of potential visual bias issues in AVS, specifically: (1) {silent object bias}, where models segment visually salient but silent objects; (2) {background noise bias}; and (3) {irrelevant sound bias}, where unrelated sounds are misattributed to visible objects.

\subsection{Evaluation Protocols}
\label{subsec:eval}
A robust AVS model should not only accurately segment sound-producing objects but also reliably suppress predictions when no valid audio-visual correspondence exists. To enable this comprehensive evaluation, we propose new metrics that assess both aspects of AVS performance.

Let $\mathcal{P}$ and $\mathcal{N}$ denote sets of positive and negative samples, respectively. For positive samples, following established protocols~\cite{zhou2022audio, gao2024avsegformer, ma2024stepping}, we employ mean Intersection over Union (mIoU) and F-score to evaluate segmentation accuracy. For negative ones, we introduce complementary metrics to capture different aspects of model robustness.

\noindent
\textit{False Positive Rate (FPR):}
\begin{equation}
\text{FPR} = \frac{\sum_{x\in \mathcal{M_{\text{pred}}} } m(x)}{H \cdot W},
\end{equation}
where $m(x)$ denotes the binary indicator (0 or 1) for pixel $x$ in the predicted mask.
FPR measures the proportion of incorrectly activated pixels in negative scenarios, directly assessing the model's tendency to generate false predictions.

To evaluate overall performance across both positive and negative cases, we propose three global metrics.

\noindent
\textit{Global mIoU (G-mIoU):} 
\begin{equation} \label{eq:gmiou}
\text{G-mIoU} = \frac{2 \cdot \text{mIoU}_\mathcal{P} \cdot (1 - \text{mIoU}_\mathcal{N})}{\text{mIoU}_\mathcal{P} + (1 - \text{mIoU}_\mathcal{N})}.
\end{equation}
where $\text{mIoU}_\mathcal{P}$ is the mIoU for positive samples, and $\text{mIoU}_\mathcal{N}$ is for negative samples. G-mIoU balances region-level accuracy, emphasizing the model's ability to maintain precise segmentation boundaries while suppressing false activations. A high score indicates accurate object delineation in positive cases and clean masks in negative cases.

\noindent
\textit{Global F-score (G-F):} 
\begin{equation} \label{eq:gf}
\text{G-F} = \frac{2 \cdot \text{F}_\mathcal{P} \cdot (1 - \text{F}_\mathcal{N})}{\text{F}_\mathcal{P} + (1 - \text{F}_\mathcal{N})}.
\end{equation}
G-F provides a pixel-level assessment that equally weighs precision and recall, which is essential for evaluating both false positives and false negatives. This metric is particularly sensitive to small errors that may be overlooked by IoU-based measures.

\noindent
\textit{Global False Positive Rate (G-FPR):}
\begin{equation} \label{eq:gfpr}
\text{G-FPR} = \frac{1}{|\mathcal{N}|} \sum_{i \in \mathcal{N}} \text{FPR}_{i}.
\end{equation}
This metric specifically focuses on false activation suppression across all negative conditions. While G-mIoU and G-F balance positive and negative performance, G-FPR provides a dedicated measure of a model's robustness against different types of audio distractors.

The combination of these metrics provides a comprehensive evaluation framework: G-mIoU captures region-level accuracy, G-F ensures pixel-level precision, and G-FPR specifically measures robustness to negative conditions. Together, they enable thorough assessment of both segmentation quality and prediction suppression capabilities.

\begin{figure*}[ht]
    \centering

        \centering
        \includegraphics[width=0.9\textwidth]{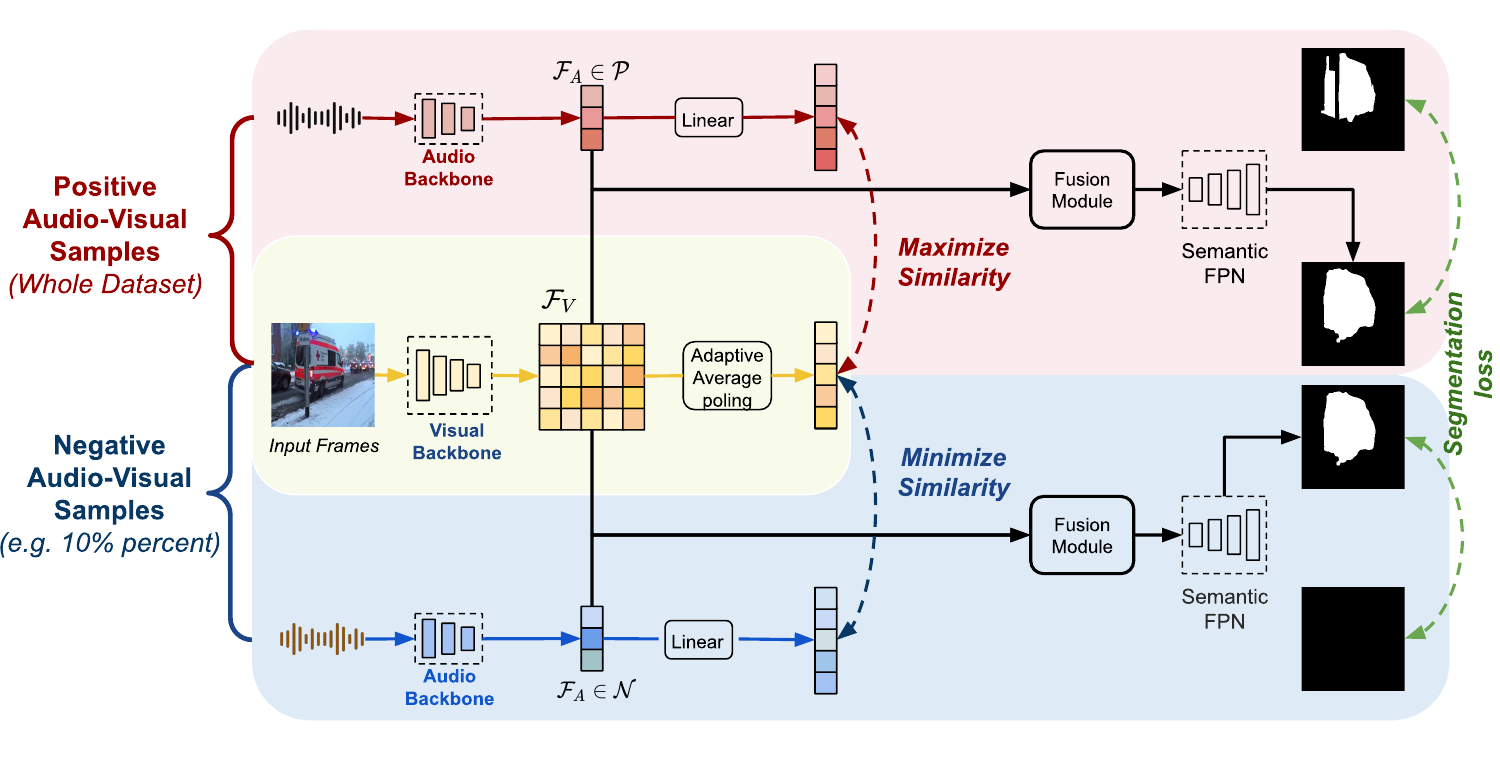}
        \vspace{-8mm}
        \caption{\textbf{Framework Overview.} Given video frames and an audio clip as inputs, our approach can robustly identify and segment sounding objects in video frames. Positive audio-visual pairs represent aligned sound sources, while negative pairs, such as silence or offscreen sounds, correspond to empty masks. The model uses separate visual and audio encoders to extract modality-specific features, applies similarity-based alignment optimized with classifier guidance in a contrastive manner, and integrates features through a fusion module. Positive pairs maximize similarity, while negative pairs minimize it, using a small portion (10\%) of the dataset for improved boundary delineation. This dual-stream design facilitates segmentation by distinguishing sound-relevant regions in complex scenes.}
        \label{fig:AVSBench}

    \vspace{-3mm}
\end{figure*}

\section{Method}
\label{sec:method}

In this section, we first present a framework overview in Sec.~\ref{subsec:framework_overview}. Upon the framework, we detail our approach to address the bias problem in AVS through three key components: balanced audio-visual pair construction (Sec. \ref{subsec:Learning_Balanced}), classifier-guided similarity learning (Sec. \ref{subsec:bce_guide} ), and joint segmentation training (Sec. \ref{subsec:total_loss}). To validate our approach, we apply it to two representative AVS models: AVSBench \cite{zhou2022audio} and AVSegFormer \cite{gao2024avsegformer}. The architecture of AVSBench is described in Sec.~\ref{subsec:preliminary}. Due to space constraints, implementation details for \cite{gao2024avsegformer} are provided in the appendix.


\subsection{Preliminary: AVS Architecture}
\label{subsec:preliminary}

\textbf{Encoder:} We employ an encoder structure that separately processes audio clip $A$ and visual frames $V$. Specifically, input audio is converted into spectrograms and processed through a VGGish-based network \cite{hershey2017vggish}, pre-trained on AudioSet \cite{gemmeke2017audioset}, to generate audio feature $\mathcal{F}_A \in \mathbb{R}^d$ where $d = 128$. For visual inputs $V$, we utilize a transformer-based backbone~\cite{wang2022pvt} to extract hierarchical visual features.$\mathcal{F}_{V_i} \in \mathbb{R}^{h_i \times w_i \times C_i}$, where $(h_i,w_i) = (H,W)/2^{i+1}, i = 1,\ldots,n$. The number of levels is set to $n = 4$ in all experiments. 

\noindent \textbf{Cross-Modal Fusion:} Following the work in \cite{zhou2022audio}, the fusion process involves an Atrous Spatial Pyramid Pooling (ASPP) module~\cite{chen2017aspp} that manipulates the visual feature maps to enhance object recognition capabilities in varying receptive fields. Subsequently, audio features are integrated to reinforce the identification of sounding objects, crucial for precise segmentation in mixed audio scenarios.

\noindent\textbf{Decoder:} The decoder leverages a Panoptic-FPN \cite{kirillov2019panoptic} architecture, which sequentially processes outputs from the fusion stage and refines them through upsampling, aiming to recover detailed segmentations at the original scale.

\noindent \textbf{Segmentation Loss:} The segmentation objective is the binary cross-entropy loss for basic segmentation accuracy.
\begin{equation} 
\mathcal{L}_{\text{Seg}} = \mathcal{L}_{BCE}(\mathcal{M}_{pred}, \mathcal{M}_{gt}),  
\end{equation}
where $\mathcal{M}_{pred}$ is the predicted segmentation mask, $\mathcal{M}_{gt}$ is the ground-truth (GT) mask.

\subsection{Framework Overview}
\label{subsec:framework_overview}

Our framework, as illustrated in Fig.~\ref{fig:AVSBench}, processes both positive and negative audio-visual pairs to learn robust correspondence for segmentation. Built upon the presented AVS architecture, our model achieves balanced training by incorporating negative audio-visual pairs, enhancing robustness in AVS. Within this framework, audio and visual features are extracted and used to compute cosine similarity scores for both positive pairs $\mathcal{P}$ and negative pairs $\mathcal{N}$, allowing the model to differentiate aligned from unaligned audio-visual pairs. For mask prediction, we employ a segmentation module that combines a fusion module and an FPN decoder, enabling precise segmentation of sound-producing objects. The dual-stream design allows the model to accurately identify sound-relevant regions in complex scenes while suppressing predictions when no valid audio-visual correspondence exists. The following sections detail each component and their integration within the framework.

\subsection{Learning with Balanced Audio-Visual Pairs}
\label{subsec:Learning_Balanced}

In real-world scenarios, audio-visual correspondence is inherently dynamic~\cite{chen2022comprehensive, chakraborty2023multimodal, yang2024combo}. A visible object may or may not be producing sound at any given moment—for instance, a person may be speaking or silent, and a car may be running or stationary. Additionally, sounds may come from off-screen sources or be ambient noise. This variability requires AVS models to learn true audio-visual association rather than assume that all visible objects are sound sources.

Existing AVS models have been trained with predominantly \textit{positive} audio-visual pairs, where audio and visual signals align, and salient objects are typically the sound sources. This encourages AVS models to rely solely on visual information, bypassing true multimodal integration.

Motivated by this insight, we propose a critical requirement: models must be trained with both positive and negative audio-visual pairs. This balanced approach ensures that the model learns not only when to segment objects that make sounds but also, crucially, when to suppress segmentation predictions for visually salient but silent objects.


Given a video clip with its corresponding audio signal, we construct two types of pairs:

\textit{Positive Pairs ($\mathcal{P}$):} Original audio-visual pairs where the audio corresponds to visual objects in the frame. These pairs represent valid correspondence cases and constitute the majority (approximately 90\%) of training samples.

\textit{Negative Pairs ($\mathcal{N}$):} We deliberately create challenging negative scenarios by:
 1) Replacing the original audio with silence;
 2) Replacing the original audio background noise or ambient sounds;
 3) Using off-screen sounds that are semantically distinct from visible objects.

We maintain a 10\% of negative pairs during training, which we empirically found to optimally balance robustness, segmentation accuracy, and training efficiency.  Expanding the diversity of training samples is anticipated to further enhance the model's robustness.


\subsection{Classifier-Guided Feature Alignment}
\label{subsec:bce_guide}
However, we observed that simply introducing negative pairs is insufficient to mitigate the visual bias, as show in Table~\ref{tab:only_negative}. Due to the inherent bias in existing models, which often fail to effectively utilize audio information, the model tends to behave more like a purely visual segmentation model. Without explicit guidance, adding negative pairs can lead to confusion during training, as the model alternates between predicting object masks and empty masks. This ultimately degrades performance, not only on the original dataset but also in negative conditions, where the model may continue to produce object masks despite the absence of valid audio-visual correspondence.

While balanced training with positive and negative pairs exposes the model to diverse scenarios, it needs explicit guidance to learn when audio and visual features truly correspond. To address this, we propose using a classifier to directly supervise audio-visual similarity learning, creating clear decision boundaries for correspondence detection.



Given multi-scale visual features $\mathcal{F}_i \in \mathbb{R}^{h_i \times w_i \times C_i}$ from the backbone, we use the final-stage features $\mathcal{F}_4 \in \mathbb{R}^{h_4\times w_4 \times C_4}$ and audio features $\mathcal{F}_A \in \mathbb{R}^{D_a}$ for similarity computation. We project $\mathcal{F}_A$ to $C_4$ dimensions via a linear layer and apply spatial pooling to $\mathcal{F}_4$ to obtain aligned features $\hat{\mathcal{F}}_A, \hat{\mathcal{F}}_V \in \mathbb{R}^{C_4}$. 
Their correspondence is then computed through cosine similarity:
\begin{equation}
   s(F_A, F_V) = \text{cos}(\hat{\mathcal{F}}_A, \hat{\mathcal{F}}_V).
\end{equation}

We then apply BCE loss to explicitly guide similarity learning in a contrastive manner:
\begin{equation}
\begin{split}
   \mathcal{L}_{\text{BCE}} = -\frac{1}{|\mathcal{P}| + |\mathcal{N}|} & \sum_{j=1}^{|\mathcal{P}| + |\mathcal{N}|}  \left( y_j \log \sigma(s_j) \right. \\
   & \left. + (1 - y_j) \log (1 - \sigma(s_j)) \right),
\end{split}
\end{equation}
where $\sigma(\cdot)$ is the sigmoid function, $y_j$ is the binary label (1 for positive pairs, 0 for negative pairs), and $|\mathcal{P}| + |\mathcal{N}|$ is the total number of positive pairs and the total number of negative pairs respectively. By explicitly supervising the similarity learning, the BCE loss forces the model to maximize similarity for positive pairs (where valid audio-visual correspondence exists) and minimize it for negative pairs (where no correspondence is present). This guidance helps the model learn to interpret audio as a cue for segmentation only when there is a meaningful alignment with the visual input, reducing confusion in cases without correspondence.

\subsection{Joint Training with Segmentation}
\label{subsec:total_loss}
Our total loss objective function $\mathcal{L}$ can be computed as follows:
\begin{equation}
   \mathcal{L} = \lambda \mathcal{L}_{\text{BCE}} + \mathcal{L}_{\text{Seg}},
\end{equation}
where $\lambda$ is a balancing weight. Together, these loss terms enforce robust and effective learning in AVS models:
1) The first term determines whether segmentation should occur based on audio-visual correspondence;
2) The second term ensures correct segmentation masks when correspondence exists;
3) For negative pairs, the empty GT masks naturally guide the segmentation loss to suppress predictions.

This simple, well-motivated approach can achieve strong performance without relying on complex model modifications, making our method easier to implement, tune, and integrate with existing AVS architectures.



\begin{table*}[!htbp]
\centering
\resizebox{\textwidth}{!}{%
\begin{tabular}{c|c|cc|ccc|ccc|ccc|ccc}
\hline
& & \multicolumn{2}{c|}{\textbf{Positive audio input}}&  \multicolumn{9}{c|}{\textbf{Negative audio input}}& \multicolumn{3}{c}{\textbf{Global metric}}\\
 & & \multicolumn{2}{c|}{}&  \multicolumn{3}{c}{\textbf{Slience}}&\multicolumn{3}{c}{\textbf{Noise}}&   \multicolumn{3}{c|}{\textbf{Offscreen sound}}& \multicolumn{3}{c}{}\\ \hline
 \textbf{Test set}&  \textbf{Model}& \textbf{mIoU ↑}& \textbf{F-score ↑}& \textbf{mIoU ↓}&  \textbf{F-score ↓}&\textbf{FPR ↓}&\textbf{mIoU ↓}& \textbf{F-score ↓}& \textbf{FPR ↓}& \textbf{mIoU ↓}&  \textbf{F-score ↓}&\textbf{FPR ↓}& \textbf{G-mIoU↑}& \textbf{G-F↑}&\textbf{G-FPR↓}\\ \hline
\multirow{7}{*}{\textbf{AVSBench-S4}}& AVSBench~\cite{zhou2022audio}& 78.7& 87.9& 76.6&   87.1&0.19&77.6& 88.0& 0.18& 78.2&  88.2&0.19
& 35.032&  21.479&0.186
\\
 & AVSegFormer~\cite{gao2024avsegformer}& 82.1& 89.9& 83.0&   90.4&0.19&83.0& 90.4& 0.19& 83.0&  90.4&0.19
& 28.199&  17.355&0.188
\\
 & Stepping-Stones~\cite{ma2024stepping}& 83.2& 91.3& 82.2&   91.3&0.19&82.2& 91.3& 0.19& 82.5&  91.3&0.19
& 28.980&  15.806&0.190
\\
 & SAMA-AVS~\cite{liu2024annofree}& 83.1& 90.0& 56.2&   69.1&0.17&59.3& 73.8& 0.13& 68.7&  79.0&
0.17& 52.688&  40.417&0.155
\\
 & CAVP~\cite{chen2024cavp}& 78.7& 88.8& 78.7&   88.8&0.19&78.7& 88.8& 0.19& 78.7&  88.8&0.19
& 33.526&  19.891&0.185\\
 & COMBO~\cite{yang2024combo}& \textbf{84.7}& \textbf{91.9}& 84.6&   91.9&0.19&84.6& 91.9& 0.19& 84.6&  91.9&
0.19& 26.062&  14.888&0.190\\
     
     & \cellcolor{lightblue} \textbf{AVSBench + Ours} & \cellcolor{lightblue} 78.1& \cellcolor{lightblue} 88.2& \cellcolor{lightblue} \textbf{0.2}& \cellcolor{lightblue} \textbf{22.6}& \cellcolor{lightblue} \textbf{0.00}& \cellcolor{lightblue} \textbf{0.2}& \cellcolor{lightblue} \textbf{22.6}& \cellcolor{lightblue} \textbf{0.00}& \cellcolor{lightblue} \textbf{0.2}& \cellcolor{lightblue} \textbf{22.6}& \cellcolor{lightblue} \textbf{0.00}& \cellcolor{lightblue} \textbf{87.672}& \cellcolor{lightblue} \textbf{82.461}& \cellcolor{lightblue} \textbf{0.000}\\
    
     & \cellcolor{lightblue} \textbf{AVSegFormer + Ours} & \cellcolor{lightblue} 74.2& \cellcolor{lightblue} 84.8& \cellcolor{lightblue} \textbf{0.2}& \cellcolor{lightblue} \textbf{22.6}& \cellcolor{lightblue} \textbf{0.00}& \cellcolor{lightblue} \textbf{0.3}& \cellcolor{lightblue} \textbf{22.7}& \cellcolor{lightblue} \textbf{0.00}& \cellcolor{lightblue} \textbf{0.5}& \cellcolor{lightblue} \textbf{22.9}& \cellcolor{lightblue} \textbf{0.00}&  \cellcolor{lightblue}\textbf{85.069}& \cellcolor{lightblue} \textbf{80.849}& \cellcolor{lightblue}\textbf{0.001}\\ \hline

\multirow{7}{*}{\textbf{AVSBench-MS3}}& AVSBench~\cite{zhou2022audio}& 54.0& 64.5& 27.6&   53.5&0.05&31.7& 57.4& 0.05& 42.2&  62.4&0.09
& 59.468&  51.036&0.072
\\
 & AVSegFormer~\cite{gao2024avsegformer}& 61.3& 73.8& 53.2&   68.2&0.13&47.5& 63.8& 0.09& 50.3&  66.0&0.11
& 54.889&  46.571&0.103
\\
 & Stepping-Stones~\cite{ma2024stepping}& 67.3& 77.6& 45.6&   
72.5&0.09&43.8& 
72.3& 0.08& 41.0&  63.3&0.15& 61.439&  43.937&0.114
\\
 & SAMA-AVS~\cite{liu2024annofree}& \textbf{68.6}& \textbf{78.3}& 29.7&   
36.8&0.09&39.2& 
46.6& 0.12& 44.1&  49.9&
0.14& 65.308&  65.038&0.125
\\
 & CAVP~\cite{chen2024cavp}& 45.8& 61.7& 45.8&   61.7&0.11&45.8& 61.7& 0.11& 45.8&  61.7&0.11
& 49.647&  47.262&0.110\\
 & COMBO \cite{yang2024combo}& 59.2& 71.2& -&   
-&-&-& -& -& -&  -&
-& -&  -&-\\ 

    & \cellcolor{lightblue} \textbf{AVSBench + Ours} & \cellcolor{lightblue} 51.3& \cellcolor{lightblue} 64.5& \cellcolor{lightblue} 9.8& \cellcolor{lightblue} 17.7& \cellcolor{lightblue} \textbf{0.00}& \cellcolor{lightblue} \textbf{9.9}& \cellcolor{lightblue} \textbf{25.8}& \cellcolor{lightblue} \textbf{0.00}& \cellcolor{lightblue} \textbf{9.1}& \cellcolor{lightblue} \textbf{20.3}& \cellcolor{lightblue} \textbf{0.00}& \cellcolor{lightblue} \textbf{65.427}& \cellcolor{lightblue} \textbf{70.911}& \cellcolor{lightblue} \textbf{0.001}\\ 

    & \cellcolor{lightblue} \textbf{AVSegFormer + Ours} & \cellcolor{lightblue} 61.5& \cellcolor{lightblue} 74.0& \cellcolor{lightblue} \textbf{9.1}& \cellcolor{lightblue} \textbf{17.0}& \cellcolor{lightblue} \textbf{0.00}& \cellcolor{lightblue} \textbf{9.4}& \cellcolor{lightblue} \textbf{17.2}& \cellcolor{lightblue} \textbf{0.00}& \cellcolor{lightblue} \textbf{9.1}& \cellcolor{lightblue} \textbf{17.0}& \cellcolor{lightblue} \textbf{0.00}& \cellcolor{lightblue} \textbf{73.354}& \cellcolor{lightblue} \textbf{78.244}& \cellcolor{lightblue} \textbf{0.000}\\ 

\hline
\end{tabular}%
}
\vspace{-2mm}
\caption{Performance comparison of various models on different audio input types and global metrics.}
\label{tab:main_table}
\end{table*}

\section{Experiment}



\begin{figure*}[h]
\vspace{-2mm}
    \centering
    \includegraphics[width=1.0\textwidth]{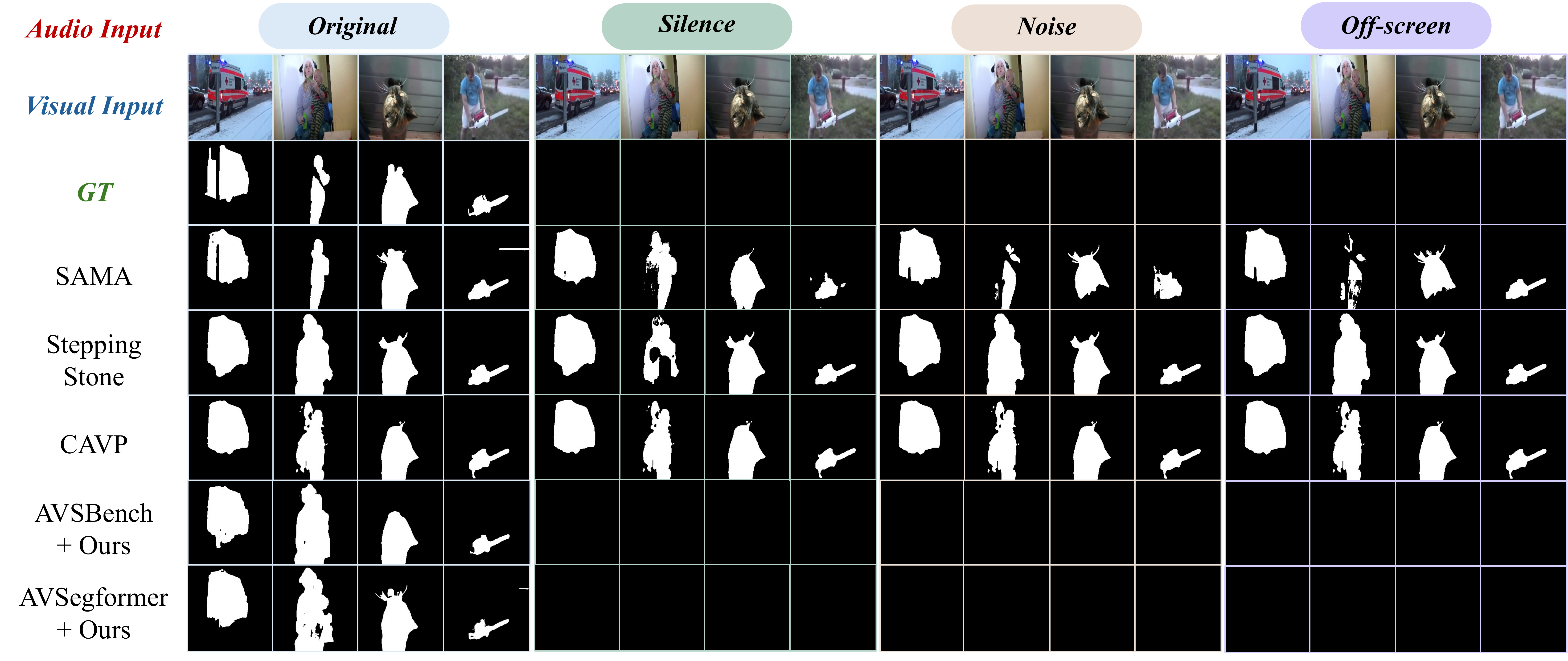}

    \caption{\textbf{Performance comparison of different AVS models under various audio conditions on Robust-S4 dataset}. Existing SOTA methods \cite{liu2024annofree, ma2024stepping, chen2024cavp} segment objects primarily based on visual salience, exhibiting a strong visual bias. In contrast, our approach achieves accurate segmentation with original audio while successfully reject predict in negative scenarios (e.g., silence, noise, off-screen).}
    \label{fig:examples}
\vspace{-4mm}
\end{figure*}


\subsection{Setup}

\noindent\textbf{Dataset.} We utilize the AVSBench-Robust Benchmark for our evaluation, which is designed to rigorously assess AVS capabilities. Further details on the dataset specifics and video categories have been discussed in Sec. \ref{sec:problem_benchmark}.

\noindent\textbf{Baselines.} We benchmark our model against notable methods including AVSBench~\cite{zhou2022audio} and AVSegFormer~\cite{gao2024avsegformer}, representing fusion-based and prompt-based approaches, respectively. We also compared our method with the CAVP~\cite{chen2024cavp}, Stepping-Stones~\cite{ma2024stepping}, SAMA-AVS~\cite{liu2024annofree} and COMBO~\cite{yang2024combo}. These baselines allow us to demonstrate the broad applicability of our method by comparing it against state-of-the-art models designed to address different aspects of audio-visual segmentation.

\noindent\textbf{Evaluation Metrics.} 
Evaluation metrics, including mIoU, F1 score, FPR, and G-mIou, G-F, G-FPR, are used to assess the segmentation accuracy and robustness of AVS models.

\noindent\textbf{Implementation:} Our implementation for the AVSBench model employ the Pyramid Vision Transformer (PVT-v2)~\cite{wang2022pvt} pretrained on the ImageNet dataset~\cite{russakovsky2015imagenet} as the visual backbone, which processes video frames of size $H \times W = 224 \times 224$ and output multiple scales visual feature $\mathcal{F}_{V_i} \in \mathbb{R}^{h_i \times w_i \times C_i}$ for $i = 1,\ldots,4$, The channel dimensions $C_i$ correspond to \{64, 128, 320, 512\} for each respective scale.
For audio input, we employ VGGish~\cite{hershey2017vggish} pretrained on AudioSet~\cite{gemmeke2017audioset} to extract features $\mathcal{F}_A \in \mathbb{R}^{128}$ from each one-second audio clips.
This model is trained using the Adam~\cite{kingma2014adam} optimizer with a learning rate of $1 \times 10^{-4}$, batch size of 4, and loss weighting factor $\lambda = 1$. Training durations are 15 epochs for the semi-supervised S4 setup and 30 epochs for the fully-supervised MS3 setup on an NVIDIA RTX A5000 GPU.

\subsection{Experimental Comparison}
Our extensive experimental comparisons reveal several significant findings in AVS performance as shown in Table~\ref{tab:main_table}. 

\noindent
\textbf{SOTA methods fail under negative audio conditions}, demonstrating a strong visual bias and ineffective audio-visual integration.
Surprisingly, recent methods like Stepping-Stones~\cite{ma2024stepping} and CAVP~\cite{chen2024cavp} achieve nearly identical mIoU and F-scores regardless of whether the input audio is silent, irrelevant, or noisy.  These methods consistently exhibit high False Positive Rates (FPR), ranging from 0.17 to 0.19 across all negative scenarios on AVSBench-S4, indicating a significant reliance on visual cues.  This issue also impacts their global metrics, with G-mIoU scores between 28.19 and 35.03. These results suggest that these methods fail to effectively leverage audio information in this multimodal segmentation task. While this issue is somewhat less pronounced on the MS3 dataset, it remains present.

\vspace{1mm}
\noindent
\textbf{Our method resolves bias while maintaining performance.} When integrated with AVSBench~\cite{zhou2022audio}, our approach performs comparable positive audio performance (mIoU: 78.1, F-score: 88.2) while achieving perfect robustness to negative audio inputs with an FPR of 0.00 across all negative conditions. Similarly, our AVSegFormer~\cite{gao2024avsegformer} variant demonstrates only minimal degradation in positive audio metrics while achieving perfect FPR scores. Most notably, our approach achieves superior global metrics, with our AVSBench variant reaching a G-mIoU of 87.672 and G-F score of 82.461, substantially outperforming existing methods. The consistent improvement in robustness across two very different AVS architectures demonstrates the effectiveness and generality of our approach. Fig.~\ref{fig:examples} provides examples of S4 dataset visualizations. Visualizations for MS3 dataset are included in the supplementary material.

\vspace{1mm}
\noindent
\textbf{Our method excels in global metrics across scenarios}, showing consistent improvements in both the single-source and more complex multi-source settings. In the MS3, it maintains perfect robustness with an FPR of 0.00 in all negative conditions and achieves impressive global metrics; the AVSegFormer variant records a G-mIoU of 73.354 and a G-F score of 78.244. These results confirm our method's scalability and its significant advancement in addressing the longstanding limitations of existing AVS methods.


\begin{figure}[t]
\vspace{-2mm}
    \centering
    \begin{subfigure}[t]{0.24\textwidth}
        \centering
        \includegraphics[width=\textwidth]{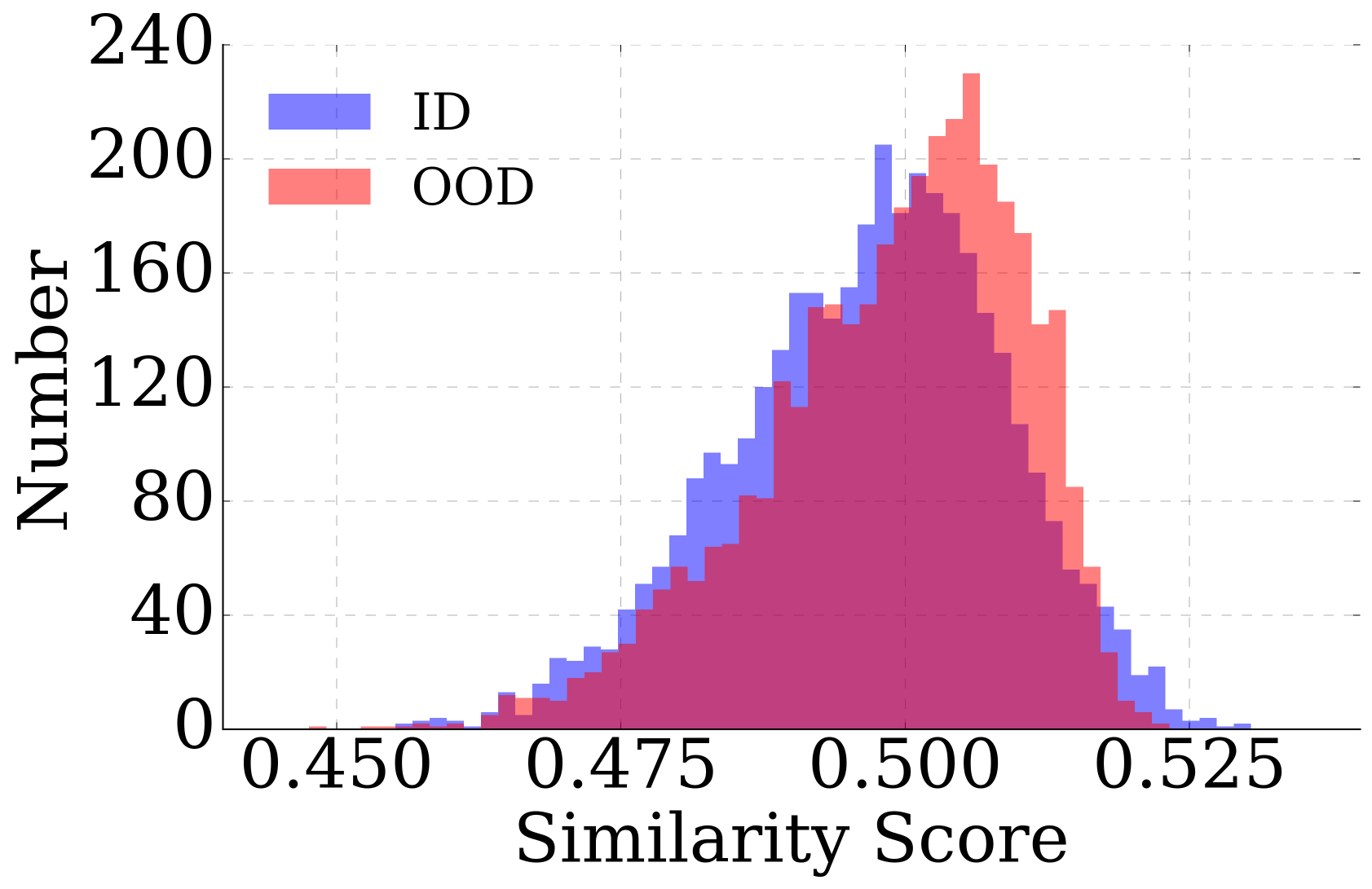}
        \caption{Before}
    \end{subfigure}%
    \begin{subfigure}[t]{0.24\textwidth}
        \centering
        \includegraphics[width=\textwidth]{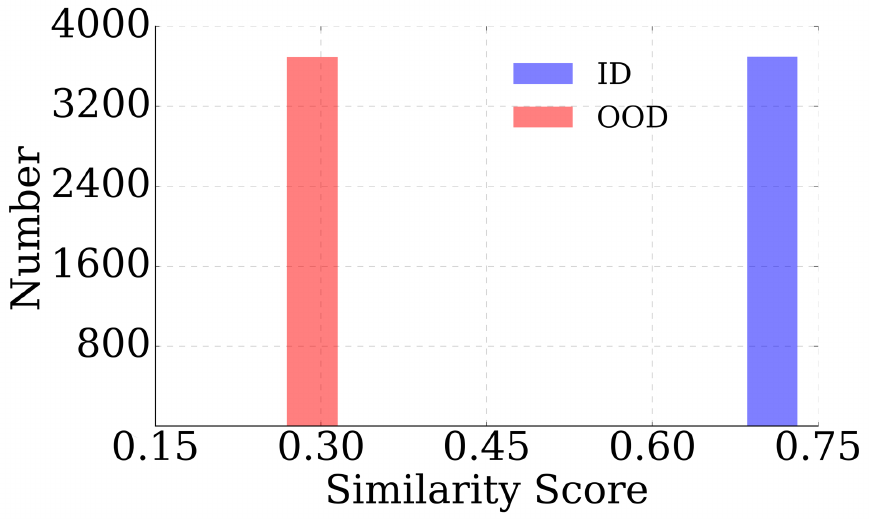}
        \caption{After}
    \end{subfigure}
    \caption{Cosine similarity distributions between paired features before and after training.(a) Positive and negative pairs exhibit similar distributions, indicating the model’s limited ability to distinguish audio-visual correspondence. (b) After training with classifier-guided similarity learning, the distributions are well-separated, demonstrating the model's enhanced capability to identify valid audio-visual pairs. }
    \label{fig:similarity_hist}
\vspace{-3mm}
\end{figure}

\subsection{Impact of Positive-Negative Pair Ratio}


Our investigation into the ratio of positive to negative audio-visual pairs reveals important insights about training data composition for robust audio-visual segmentation.

As illustrated in Table~\ref{tab:pair_ratio_study}, we found that \textit{introducing negative samples, even in small proportions, dramatically improves performance.} Without negative samples, the baseline model shows poor performance with the G-mIoU of 35.032 and a high FPR of 0.186 on the S4 dataset.
Introducing just 10\% negative samples yields substantial gains, improving G-mIoU to 87.672 and reducing FPR to 0.000. Similar improvements are observed in the MS3 dataset, where G-mIoU increases from 59.468 to 65.427 and FPR drops from 0.072 to 0.001, demonstrating the crucial role of negative samples in developing robust models.

While further increasing the proportion of negative samples (to 20\% or 30\%) maintains similar performance levels, the marginal gains are minimal compared to the 10\% setting. For instance, the G-mIoU difference between 10\% and 20\% negative samples is less than 0.3\% on S4 and 2.5\% on MS3. Considering that adding 10\% negative pairs only increases training time by approximately 10\% while achieving nearly optimal performance, we adopt this ratio as our default setting, offering an efficient balance between robustness and training cost.

\subsection{Abaltion Study}



A common assumption might be that simply adding negative samples would enhance the model's ability to distinguish between sound-producing and non-sound-producing visual regions in AVS. To test this hypothesis, we conducted an ablation study using AVSBench~\cite{zhou2022audio} as the baseline, comparing configurations with and without negative samples and classifier guidance, as summarized in Table~\ref{tab:only_negative}.
\noindent
\textbf{Negative samples alone fail to address the bias problem.} 
Our experimental results highlight the limitations of using only negative samples. Without explicit loss guidance, adding negative pairs not only fails to improve performance but can even lead to significant degradation, particularly on the challenging MS3 dataset. Here, G-mIoU drops from 59.47 to 55.49, and G-F decreases dramatically from 51.04 to 30.06. This degradation occurs because the model becomes confused when alternately exposed to scenarios requiring empty predictions and those with salient object mask predictions, ultimately compromising performance even on standard positive audio inputs.


\noindent
\textbf{Combining negative samples with classifier guidance enables robust segmentation.}  Our full approach shows substantial improvements across all metrics. On the Robust S4 dataset, we achieve G-mIoU of 87.672, G-F of 82.461, and perfect G-FPR of 0. Similar gains are observed on MS3, with G-mIoU of 66.605, G-F of 70.590, and near-perfect G-FPR of 0.004. 
The effectiveness of classifier guidance is illustrated in Fig. \ref{fig:similarity_hist}: initially, audio-visual feature similarities cluster around 0.5 for both positive and negative pairs; after training, they are well-separated (0.75 for positive vs. 0.30 for negative), demonstrating enhanced discrimination. This improved feature alignment enables strong  performance on positive cases while accurately suppressing predictions in negative scenarios.
The classifier guidance serves as a critical learning framework for effectively utilize negative samples while maintaining its original capabilities, resulting in a robust AVS system. 

Further experiments on unseen audio categories demonstrate the generalization capability of our approach. Due to space constraints, we refer readers to the supplementary material for detailed results.

\begin{table}[t]
\vspace{-2mm} 
\centering
\resizebox{\columnwidth}{!}{
\begin{tabular}{c|c|c|c|c|c|c}
\hline
 & Model & Pos Pairs & Neg Pairs & G-mIoU & G-F & G-FPR \\
\hline
\multirow{4}{*}{S4} & Baseline & 100\% & 0\% & 35.032 & 21.479 & 0.186 \\
\cline{2-7}
& \multirow{3}{*}{Ours} & 90\% & 10\% & \underline{87.672} & \underline{82.461} & 0.000 \\
& & 80\% & 20\% & \textbf{87.780} & \textbf{82.114} & 0.000 \\
& & 70\% & 30\% & 87.204 & 82.233 & 0.000 \\
\hline
\multirow{4}{*}{MS3} & Baseline & 100\% & 0\% & 59.468 & 51.036 & 0.072 \\
\cline{2-7}
& \multirow{3}{*}{Ours} & 90\% & 10\% & \underline{65.427} & \underline{70.911} & 0.001 \\
& & 80\% & 20\% & \textbf{67.909} & \textbf{72.572} & 0.003 \\
& & 70\% & 30\% & 66.251 & 72.908 & 0.000 \\
\hline
\end{tabular}
}
\vspace{-2mm} 
\caption{Impact of positive-negative ratio on AVS performance}
\label{tab:pair_ratio_study}
\end{table}

\begin{table}[t]
\centering
\resizebox{\columnwidth}{!}{  
\begin{tabular}{c|c|c|c|c|c}
\hline
 & Negative samples & $\mathcal{L}_{\text{BCE}}$& G-mIoU↑ & G-F↑ & G-FPR↓ \\
\hline
 \multirow{3}{*}{S4} & \xmark& \xmark& 35.032& 21.479& 0.186
\\
 & \cmark& \xmark& 34.847& 21.993& 0.189\\
 & \cmark& \cmark& \textbf{87.672}& \textbf{82.461}& \textbf{0.000}
\\ \hline
 \multirow{3}{*}{MS3} & \xmark& \xmark& 59.468& 51.036& 0.072
\\
 & \cmark& \xmark& 55.489& 30.057& 0.095\\
 & \cmark& \cmark& \textbf{66.605}& \textbf{70.590}& \textbf{0.004}\\
\hline
\end{tabular}
}
\vspace{-2mm} 
\caption{Effects of negative samples and classifier guidance.}
\label{tab:only_negative}
\vspace{-3mm} 
\end{table}

\section{Conclusion and Discussion}
Our comprehensive study using AVSBench-Robust reveals  that current SOTA methods exhibit strong visual bias, generating segmentation masks based predominantly on visual salience regardless of audio context. To address this issue, we introduce a simple yet effective approach combining balanced training with negative audio-visual pairs and classifier-guided feature alignment, which significantly improves model robustness while maintaining competitive performance on standard AVS tasks.
While our method effectively addresses the robustness issue, several challenges remain. Our approach is constrained by the baseline model's performance on positive samples, and real-world applications may encounter even more challenging conditions than those covered in our benchmark. 
We hope our work could inspire further research in this significant and worthwhile field.

{
    \small
    \bibliographystyle{ieeenat_fullname}
    \bibliography{main}
}

\section*{Supplementary Material}

\subsection*{A. Additional Experimental Results}

\subsubsection*{A.1 Evaluation on Unseen Audio Categories}
    

To assess generalization capability, we evaluate model performance on four diverse unseen audio (tuning fork, rooster, sheep, and thunderstorm) across both S4 and MS3 datasets.

Tables~\ref{tab:unseen_audio_s4} and~\ref{tab:unseen_audio_ms3} present comparative results between the baseline AVSBench and our approach. On S4, the baseline exhibits significant visual bias, maintaining consistently high performance (mIoU: $\sim$78\%, F-score: $\sim$88\%) regardless of audio input. Our approach effectively mitigates this bias, reducing mIoU to near-zero and decreasing false positive ratios from 0.187 to 10\textsuperscript{-5}.

The improvement is equally pronounced in the more challenging multi-source scenario (MS3). Our method effectively suppresses false predictions, achieving very low mIoU ($\sim 0.09$) and F-scores ($\sim 0.17$) across all unseen categories. For certain categories (e.g., rooster), our approach achieves complete suppression with zero false positives, demonstrating robust audio-visual correspondence even in complex multi-source scenarios.

\subsubsection*{A.2 Multi-Source (MS3) Dataset Visualizations}

\begin{figure*}[t]
\vspace{-2mm}
    \centering
    \includegraphics[width=1.0\textwidth]{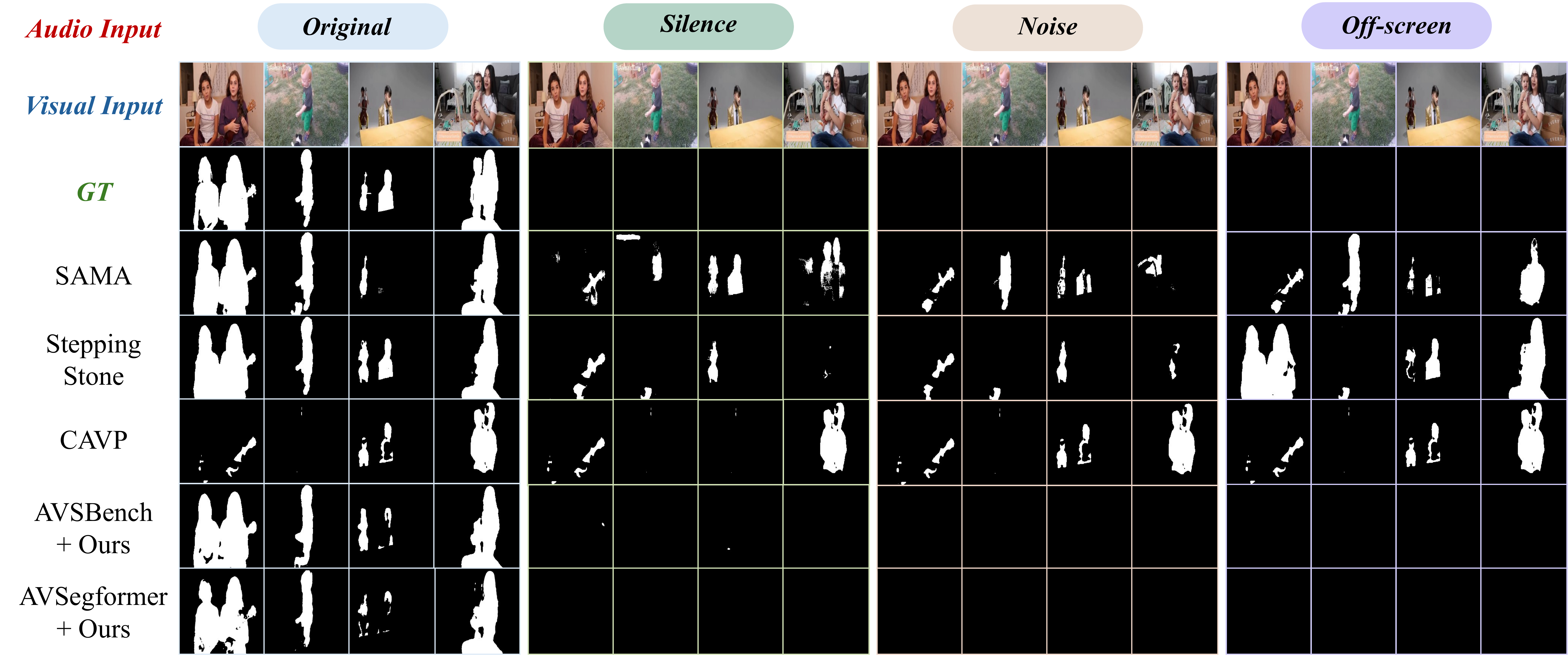}

    \caption{Performance comparison of different AVS models under various audio conditions on Robust-MS3 dataset. Existing SOTA methods \cite{liu2024annofree, ma2024stepping, chen2024cavp} segment objects primarily based on visual salience, exhibiting a strong visual bias. In contrast, our approach achieves accurate segmentation with original audio while successfully reject predict in negative scenarios (e.g., silence, noise, off-screen).}
    \label{fig:ms3_vis}
\end{figure*}

Figure~\ref{fig:ms3_vis} illustrates our method's performance on complex multi-source scenarios. The visualizations demonstrate segmentation results under various audio conditions: original audio (positive), silence, noise, and off-screen sounds, providing qualitative evidence of our model's effectiveness in handling diverse acoustic environments.

\begin{table}[t]
    \footnotesize
    \centering
    \begin{tabular}{l|c|c|c|c}
        \hline
        Category & Method & mIoU $\downarrow$ & F-score $\downarrow$ & FPR $\downarrow$ \\
        \hline
        \multirow{2}{*}{Tuning Fork} & AVSBench & 78.14 & 88.16 & 0.187 \\
        & AVSBench $+$ Ours & \textbf{0.003} & \textbf{0.248} & \textbf{0.0001} \\
        \hline
        \multirow{2}{*}{Rooster} & AVSBench & 78.19 & 88.26 & 0.186 \\
        & AVSBench $+$ Ours & \textbf{0.002} & \textbf{0.234} & \textbf{3.82e-5} \\
        \hline
        \multirow{2}{*}{Sheep} & AVSBench & 78.26 & 88.25 & 0.187 \\
        & AVSBench $+$ Ours & \textbf{0.002} & \textbf{0.236} & \textbf{8.43e-5} \\
        \hline
        \multirow{2}{*}{Thunder} & AVSBench & 78.19 & 88.23 & 0.187 \\
        & AVSBench $+$ Ours & \textbf{0.002} & \textbf{0.233} & \textbf{4.03e-5} \\
        \hline
    \end{tabular}
    \caption{Performance comparison on unseen audio categories for single-source (S4) dataset. Lower values indicate better ability to avoid false predictions for unfamiliar sounds.}
    \label{tab:unseen_audio_s4}    
\end{table}

\begin{table}[t]
    \footnotesize
    \centering
    \begin{tabular}{l|c|c|c|c}
        \hline
        Category & Method & mIoU $\downarrow$ & F-score $\downarrow$ & FPR $\downarrow$ \\
        \hline
        \multirow{2}{*}{Tuning Fork} & AVSBench & 41.44 & 62.19 & 0.091 \\
        & AVSBench $+$ Ours & \textbf{0.103} & \textbf{0.186} & \textbf{0.007} \\
        \hline
        \multirow{2}{*}{Rooster} & AVSBench & 44.52 & 63.04 & 0.102 \\
        & AVSBench $+$ Ours & \textbf{0.091} & \textbf{0.169} & \textbf{0.000} \\
        \hline
        \multirow{2}{*}{Sheep} & AVSBench & 40.14 & 62.13 & 0.082 \\
        & AVSBench $+$ Ours & \textbf{0.091} & \textbf{0.179} & \textbf{0.0001} \\
        \hline
        \multirow{2}{*}{Thunder} & AVSBench & 38.16 & 61.82 & 0.074 \\
        & AVSBench $+$ Ours & \textbf{0.091} & \textbf{0.178} & \textbf{9.07e-5} \\
        \hline
    \end{tabular}
    \caption{Performance comparison on unseen audio categories for multi-source (MS3) dataset. Lower values indicate better ability to avoid false predictions for unfamiliar sounds.}
    \label{tab:unseen_audio_ms3}    
\end{table}

\subsubsection*{A.3 Effect of feature alignment strategies}

To thoroughly evaluate our choice of cosine similarity for feature alignment, we compare it with two intuitive alternatives: Euclidean distance, which directly measures feature space proximity, and concatenation-based alignment, which preserves complete feature information. Table~\ref{table:feature_alignment} presents comparative results across both S4 and MS3 datasets.

\begin{table*}[htbp]
\centering
\resizebox{\textwidth}{!}{%
\begin{tabular}{c|c|c|c|c|c|c|c|c|c|c|c|c|c|c|c}
\hline
& \multirow{3}{*}{\begin{tabular}[c]{@{}c@{}}Guide Method\end{tabular}} & \multicolumn{2}{c|}{\textbf{Positive audio input}} & \multicolumn{9}{c|}{\textbf{Negative audio input}} & \multicolumn{3}{c}{\textbf{Global metric}} \\
\cline{5-13}
& & \multicolumn{2}{c|}{} & \multicolumn{3}{c|}{\textbf{Silence}} & \multicolumn{3}{c|}{\textbf{Noise}} & \multicolumn{3}{c|}{\textbf{Offscreen sound}} & \multicolumn{3}{c}{} \\
\cline{3-16}
\textbf{Test set} & & \textbf{mIoU↑} & \textbf{F-score↑} & \textbf{mIoU↓} & \textbf{F-score↓} & \textbf{FPR↓} & \textbf{mIoU↓} & \textbf{F-score↓} & \textbf{FPR↓} & \textbf{mIoU↓} & \textbf{F-score↓} & \textbf{FPR↓} & \textbf{G-mIoU↑} & \textbf{G-F↑} & \textbf{G-FPR↓} \\
\hline
\multirow{3}{*}{S4} & cosine & \textbf{78.1} & \textbf{88.2} & \textbf{0.2} & \textbf{22.6} & \textbf{0.000} & \textbf{0.2} & \textbf{22.6} & \textbf{0.000} & \textbf{0.2} & \textbf{22.6} & \textbf{0.00} & \textbf{87.672} & \textbf{82.461} & \textbf{0.000} \\
& Euclidean& 69.3& 82.5& 5.9& 33.8& 0.032& 0.2& 23.8& 0.000& 0.2& 24.3& 0.000& 81.144& 77.283& 0.004\\
& Concat& 77.5& 87.5& 1.0& 23.1& 0.003& 0.2& 22.6& 0.000& 0.2& 22.6& 0.000& 87.139& 82.047& 0.000\\
\hline
\multirow{3}{*}{MS3} & cosine 
& \textbf{51.3} & \textbf{64.5} & \textbf{9.8} & \textbf{17.7} & \textbf{0.00} & \textbf{9.9} & \textbf{25.8} & \textbf{0.000} & \textbf{9.1} & \textbf{20.3} & \textbf{0.000} & \textbf{66.605} & \textbf{70.590} & \textbf{0.004} \\
& Euclidean
& 43.5& 56.4& 10.7& 21.8& 0.011& 11.1& 24.2& 0.028& 9.5& 28.3& 0.001& 58.526& 64.447& 0.014\\
& Concat& 49.7& 62.8& 12.1& 21.0& 0.006& 14.8& 32.3& 0.012& 14.1& 29.0& 0.011& 63.053& 67.333& 0.011\\
\hline
\end{tabular}
}
\caption{Comparison of different feature alignment strategies. Results show performance across positive and negative audio scenarios, as well as global metrics.}
\label{table:feature_alignment}
\end{table*}

All three methods demonstrate effectiveness in audio-visual feature alignment, with each achieving reasonable performance on positive samples. However, cosine similarity exhibits superior performance, particularly in negative suppression scenarios. On S4, while concatenation-based alignment maintains competitive positive metrics (mIoU: 77.5\%), cosine similarity achieves better balance between positive performance (mIoU: 78.1\%) and negative suppression (FPR: 0.00). This pattern extends to the more challenging MS3 dataset, where cosine similarity shows notably better global metrics (G-mIoU: 66.61, G-F: 70.59) compared to both alternatives.

The advantage of cosine similarity likely stems from its inherent normalization property and focus on directional relationships, making it particularly suitable for cross-modal feature comparison. While Euclidean distance is sensitive to feature magnitude variations and concatenation may preserve redundant information, cosine similarity captures the essential semantic alignment between modalities while maintaining computational efficiency.

\noindent\textbf{Implementation Details:} For all methods, audio features are first projected from 128 to 256 dimensions to match the visual feature dimension, and visual features undergo adaptive average pooling to obtain global representations. The methods then differ in their alignment computation:
\begin{itemize}
\item \textbf{Cosine Similarity} computes normalized directional alignment using the standard cosine similarity function: $\text{similarity} = \cos(\hat{\mathcal{F}}_A, \hat{\mathcal{F}}_V)$, where $\hat{\mathcal{F}}_A$ and $\hat{\mathcal{F}}_V$ are the projected audio and visual features respectively.
\item \textbf{Euclidean Distance} measures direct feature space proximity through L2 norm: $\text{similarity} = -||\hat{\mathcal{F}}_A - \hat{\mathcal{F}}_V||_2$. The negative sign converts distance to similarity, ensuring larger values indicate stronger alignment.
\item \textbf{Concatenation-based} alignment employs a three-layer MLP that processes the concatenated features $[\hat{\mathcal{F}}_A; \hat{\mathcal{F}}_V]$ (512 dimensions). The network progressively reduces dimensionality (512 → 256 → 128 → 1) with ReLU activation and dropout (0.1) for regularization, learning more complex non-linear relationships between modalities.
\end{itemize}

\subsubsection*{A.4 Sensitivity to hyperparameter choices}

We evaluate the model's sensitivity to the classifier guidance weight $\lambda_{\text{BCE}}$ by varying its value from 0.2 to 1.0. Table~\ref{table:lambda_sensitivity} presents the quantitative results across both S4 and MS3 datasets.

\begin{table*}[t]
    \centering
    \resizebox{\linewidth}{!}{
    \begin{tabular}{c|c|cc|ccc|ccc|ccc|ccc}
        \hline
        Dataset & $\lambda_{\text{BCE}}$ & \multicolumn{2}{c|}{Positive} & \multicolumn{3}{c|}{Silence} & \multicolumn{3}{c|}{Noise} & \multicolumn{3}{c|}{Offscreen} & \multicolumn{3}{c}{Global} \\
        \cline{3-16}
        & & mIoU$\uparrow$ & F$\uparrow$ & mIoU$\downarrow$ & F$\downarrow$ & FPR$\downarrow$ & mIoU$\downarrow$ & F$\downarrow$ & FPR$\downarrow$ & mIoU$\downarrow$ & F$\downarrow$ & FPR$\downarrow$ & G-mIoU$\uparrow$ & G-F$\uparrow$ & G-FPR$\downarrow$ \\
        \hline
        \multirow{5}{*}{S4} & 1.0 & 78.15 & 88.22 & 0.16 & 22.59 & 0.00 & 0.16 & 22.59 & 0.00 & 0.16 & 22.59 & 0.00 & 87.67 & 82.46 & 0.000 \\
        & 0.8 & 78.97 & 88.78 & 0.16 & 22.60 & 0.00 & 0.16 & 22.60 & 0.00 & 0.16 & 22.60 & 0.00 & 88.18 & 82.70 & 0.000 \\
        & 0.6 & 77.98 & 88.00 & 0.18 & 22.64 & 0.00 & 0.17 & 22.63 & 0.00 & 0.17 & 22.64 & 0.00 & 87.56 & 82.34 & 0.000 \\
        & 0.4 & 78.76 & 88.56 & 1.06 & 32.21 & 0.00 & 0.16 & 22.59 & 0.00 & 0.16 & 22.65 & 0.00 & 87.54 & 80.73 & 0.000 \\
        & 0.2 & 78.95 & 88.68 & 0.22 & 22.93 & 0.00 & 0.17 & 22.85 & 0.00 & 0.18 & 22.85 & 0.00 & 88.16 & 82.50 & 0.000 \\
        \hline
        \multirow{5}{*}{MS3} & 1.0 & 51.27 & 64.51 & 9.84 & 17.72 & 0.00 & 9.93 & 25.78 & 0.00 & 9.11 & 20.33 & 0.00 & 65.43 & 70.91 & 0.001 \\
        & 0.8 & 53.32 & 66.32 & 9.64 & 18.09 & 0.00 & 11.75 & 35.07 & 0.01 & 9.81 & 31.23 & 0.00 & 66.86 & 68.98 & 0.004 \\
        & 0.6 & 51.23 & 64.94 & 13.37 & 21.38 & 0.03 & 15.20 & 37.80 & 0.02 & 9.67 & 26.79 & 0.00 & 64.55 & 67.99 & 0.011 \\
        & 0.4 & 52.75 & 65.85 & 13.66 & 26.41 & 0.02 & 10.33 & 35.82 & 0.01 & 10.33 & 35.82 & 0.01 & 66.12 & 66.57 & 0.007 \\
        & 0.2 & 52.81 & 65.56 & 9.81 & 18.01 & 0.00 & 12.99 & 35.02 & 0.01 & 9.81 & 21.94 & 0.00 & 66.32 & 69.97 & 0.005 \\
        \hline
    \end{tabular}
    }
    \caption{Performance analysis with different $\lambda_{\text{BCE}}$ values. Results demonstrate the model's robustness to this hyperparameter choice.}    
    \label{table:lambda_sensitivity}

\end{table*}

Experimental results demonstrate strong robustness to this hyperparameter choice. On S4, the G-mIoU variation remains minimal (87.54-88.18\%, $\Delta$=0.64\%), with consistent perfect negative suppression (FPR: 0.000) across all settings. The MS3 dataset shows slightly larger but still modest variations (G-mIoU: 64.55-66.86\%, $\Delta$=2.31\%), attributable to its inherently more complex multi-source scenarios. This stability suggests that the model's performance is not heavily dependent on precise hyperparameter tuning.

\subsubsection*{A.5 Detailed Analysis of the Effect of Negative Samples and Classifier Guidance}
To dissect the individual contributions of our key components, we conduct ablation experiments on both negative sample integration and classifier guidance. Table~\ref{table:full_comparison}, compare to Table 4 presents comprehensive results across different configurations.
Adding negative samples alone proves insufficient and can even degrade performance. On S4, while the baseline achieves 78.7\% mIoU with high false positives (FPR: ~0.19), incorporating only negative samples marginally improves robustness but compromises positive performance (mIoU: 79.0\%). This effect is more pronounced on MS3, where positive performance significantly degrades (mIoU drops from 54.0\% to 51.6\%) while maintaining high false positive rates. This degradation occurs because the model, without explicit guidance, struggles to establish clear decision boundaries between valid and invalid audio-visual correspondences.
The integration of classifier guidance ($\mathcal{L}_{\text{BCE}}$) with negative samples yields substantial improvements. This combination achieves 78.1\% mIoU on S4 while reducing FPR to nearly zero across all negative scenarios. Similarly on MS3, it maintains competitive positive performance (51.3\% mIoU) while effectively suppressing false predictions (FPR: 0.004).

\begin{table*}[htbp]
\centering
\resizebox{\textwidth}{!}{%
\begin{tabular}{c|c|c|c|c|c|c|c|c|c|c|c|c|c|c|c|c}
\hline
& \multirow{3}{*}{\begin{tabular}[c]{@{}c@{}}Negative\\samples\end{tabular}} & \multirow{3}{*}{$\mathcal{L}_{\text{BCE}}$} & \multicolumn{2}{c|}{\textbf{Positive audio input}} & \multicolumn{9}{c|}{\textbf{Negative audio input}} & \multicolumn{3}{c}{\textbf{Global metric}} \\
\cline{6-14}
& & & \multicolumn{2}{c|}{} & \multicolumn{3}{c|}{\textbf{Silence}} & \multicolumn{3}{c|}{\textbf{Noise}} & \multicolumn{3}{c|}{\textbf{Offscreen sound}} & \multicolumn{3}{c}{} \\
\cline{4-17}
\textbf{Test set} & & & \textbf{mIoU↑} & \textbf{F-score↑} & \textbf{mIoU↓} & \textbf{F-score↓} & \textbf{FPR↓} & \textbf{mIoU↓} & \textbf{F-score↓} & \textbf{FPR↓} & \textbf{mIoU↓} & \textbf{F-score↓} & \textbf{FPR↓} & \textbf{G-mIoU↑} & \textbf{G-F↑} & \textbf{G-FPR↓} \\
\hline
\multirow{3}{*}{S4} & \xmark & \xmark & 78.7 & 87.9 & 76.6 & 87.1 & 0.19 & 77.6 & 88.0 & 0.18 & 78.2 & 88.2 & 0.19 & 35.032 & 21.479 & 0.186 \\
& \cmark & \xmark & 79.0 & 88.4 & 76.6 & 86.6 & 0.19 & 77.9 & 87.7 & 0.20 & 78.5 & 88.0 & 0.19 & 34.847 & 21.993 & 0.189 \\
& \cmark & \cmark & 78.1 & 88.2 & 0.2 & 22.6 & 0.00 & 0.2 & 22.6 & 0.00 & 0.2 & 22.6 & 0.00 & 87.672 & 82.461 & 0.000 \\
\hline
\multirow{3}{*}{MS3} & \xmark & \xmark & 54.0 & 64.5 & 27.6 & 53.5 & 0.05 & 31.7 & 57.4 & 0.05 & 42.2 & 62.4 & 0.09 & 59.468 & 51.036 & 0.072 \\
& \cmark & \xmark & 51.6 & 23.3 & 34.4 & 52.1 & 0.10 & 40.2 & 58.6 & 0.10 & 45.2 & 62.3 & 0.10 & 55.489 & 30.057 & 0.095 \\
& \cmark & \cmark & 51.3 & 64.5 & 9.8 & 17.7 & 0.00 & 9.9 & 25.8 & 0.00 & 9.1 & 20.3 & 0.00 & 66.605 & 70.590 & 0.004 \\
\hline
\end{tabular}
}
\caption{Ablation study of negative samples and classifier guidance. Results show performance on positive and negative audio inputs, as well as global metrics. The checkmarks (\cmark) and crosses (\xmark) indicate whether negative samples and classifier guidance ($\mathcal{L}_{\text{BCE}}$) are used in each configuration}
\label{table:full_comparison}
\end{table*}

\subsection*{B Detailed Analysis of Audio Conditions}

\begin{figure}[t]
    \centering
    \includegraphics[width=\linewidth]{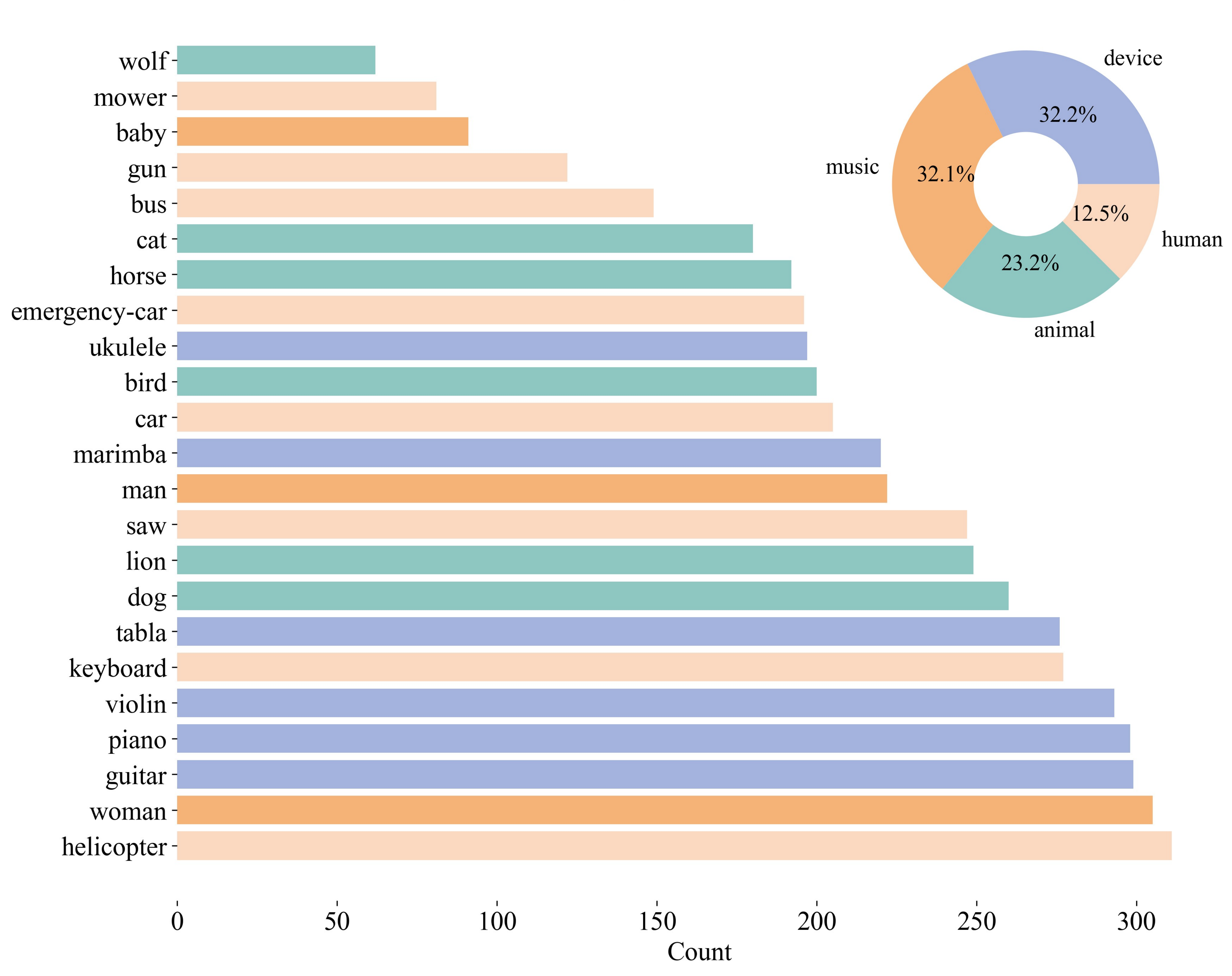}
    \caption{Distribution of cross-category combinations in multi-source scenarios. This analysis reveals common co-occurrence patterns, with music-human and animal-human being the most frequent combinations.}
    \label{fig:category_dist}
\end{figure}

\noindent\textbf{S4 Dataset Composition:} The Robust S4 dataset spans four major categories with diverse audio-visual characteristics, as shown in Figure~\ref{fig:category_dist}. The distribution is relatively balanced across device (32.2\%), music (32.1\%), and animal (23.2\%) categories, with human sounds comprising 12.5\% of the dataset. This balanced distribution helps ensure robust evaluation across different types of audio-visual scenarios.

\begin{figure}[t]
    \centering
    \includegraphics[width=\linewidth]{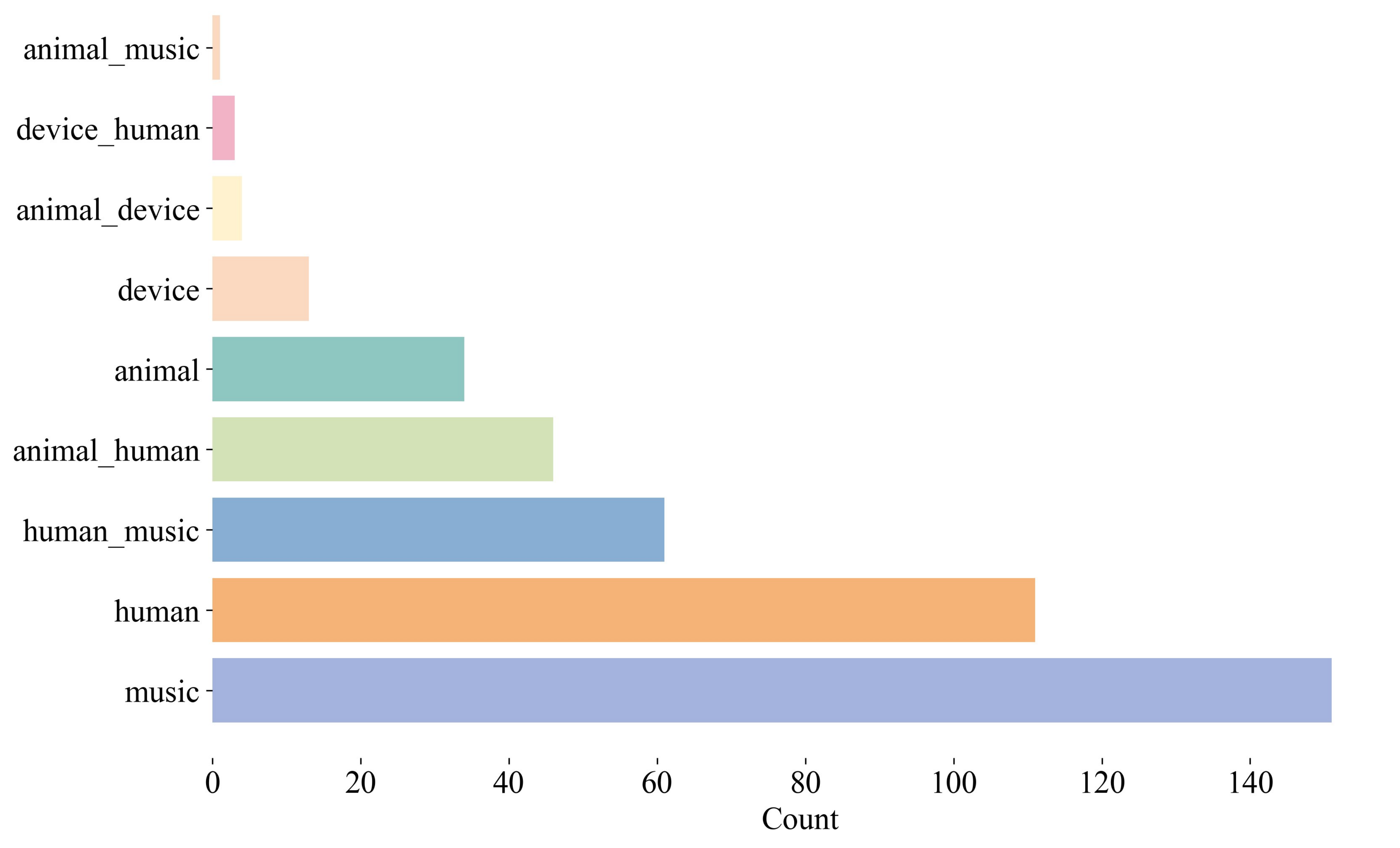}
    \caption{Distribution of multi-source audio category combinations in our dataset. The horizontal bars show the frequency of different category pairs including individual categories (music, human, animal, device) and their combinations (e.g., human\_music, animal\_human).}
    \label{fig:cross_category}
\end{figure}

\noindent\textbf{MS3 Cross-Category Analysis:} For multi-source scenarios (Figure~\ref{fig:cross_category}), we analyze the distribution of audio category combinations in our dataset. Music and human categories have the highest individual counts, followed by animal and device categories. Among cross-category combinations, human\_music shows the highest frequency, followed by animal\_human. Other combinations such as device\_human and animal\_device occur less frequently in our dataset.

\subsection*{C. Failure Case Analysis}

Our method inherits certain limitations from the base segmentation model. Specifically, in cases where the underlying model fails to correctly segment the target object, our approach will also produce incorrect results. This cascading failure occurs because our method builds upon and depends on the initial segmentation output.

For example, when the base model misidentifies object boundaries or fails to detect the target object entirely, our method cannot compensate for these fundamental segmentation errors.

Future work could explore ways to make our approach more robust to initial segmentation errors, possibly through the incorporation of additional verification mechanisms or multi-model ensemble approaches.


\subsection*{D. AVSegFormer Implementation Details}
\subsubsection*{D.1 Preliminary: AVSegFormer Architecture}
\label{subsec:avSegformer}

\begin{figure}[t]
    \centering
    \includegraphics[width=0.5\textwidth]{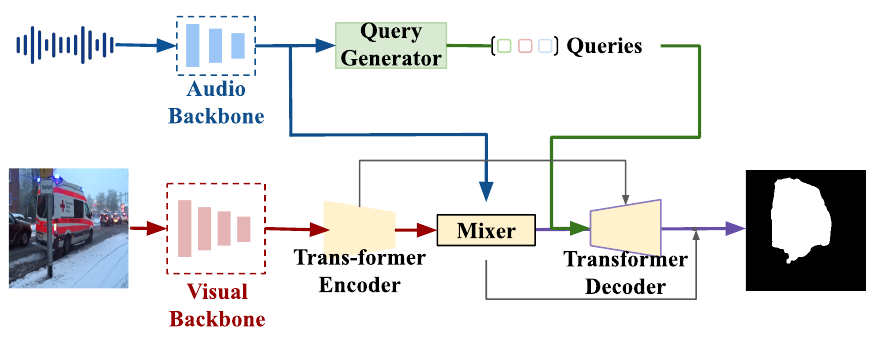}
    \caption{Overview of AVSegformer architecture. The framework processes audio and visual inputs through parallel backbone networks, generates audio-conditioned queries, and employs transformer-based encoder-decoder architecture with a specialized mixer for audio-visual feature fusion.}
    \label{fig:avsegformer_arch}
\end{figure}

AVSegFormer~\cite{gao2024avsegformer} introduces several key architectural innovations while maintaining some fundamental components from AVSBench~\cite{zhou2022audio}. As illustrated in Figure~\ref{fig:avsegformer_arch}, the framework consists of five main components: audio-visual backbone encoders, a query generator that produces audio-conditioned queries, a transformer encoder for multi-scale feature processing, an audio-visual mixer for cross-modal feature fusion, and a transformer decoder for final mask generation. The query-based design enables the model to adaptively focus on audio-relevant regions in the visual frame. Here we detail its implementation.

\noindent\textbf{Encoder:} The encoding pathway remains consistent with AVSBench, employing VGGish~\cite{hershey2017vggish} to generate audio features $\mathcal{F}_A \in \mathbb{R}^d$ ($d = 128$). For visual processing, we utilizes Pyramid Vision Transformer~\cite{wang2022pvt} to extract hierarchical features $\mathcal{F}_{V_i}$, where $i \in [1,4]$ denotes the multi-resolution stage.

\noindent\textbf{Query Generator:} A key innovation in AVSegFormer~\cite{gao2024avsegformer} is its query-based architecture. The generator processes:
\begin{itemize}
    \item Initial query: $\mathcal{Q}_{\text{init}} \in \mathbb{R}^{T\times N_{\text{query}}\times D}$
    \item Audio feature: $\mathcal{F}_{\text{audio}} \in \mathbb{R}^{T\times D}$
    \item Learnable query: $\mathcal{Q}_{\text{learn}}$
\end{itemize}
Through cross-attention mechanisms, these components are transformed into mixed queries $\mathcal{Q}_{\text{mixed}}$, which help the model adaptively focus on audio-relevant regions in the visual frame. The inclusion of learnable queries enhances the model's capability to handle various audio-visual scenarios and dataset-specific characteristics.

\noindent\textbf{Transformer Encoder:} The transformer encoder processes visual features at three different resolutions (1/8, 1/16, and 1/32 of the original size). These multi-scale features are first flattened and concatenated to form a unified representation. The concatenated features are then processed through transformer layers, after which they are reshaped back to their original spatial dimensions. The 1/8-scale features are specifically upsampled by a factor of 2 and combined with the 1/4-scale features from the visual backbone through addition, producing the final mask features $\mathcal{F}_{\text{mask}}$ at 1/4 resolution. This multi-scale processing ensures the model captures both fine details and broader contextual information.

\noindent\textbf{Audio-Visual Mixer:} The mixer implements channel attention mechanism through:
\begin{equation}
    \omega = \text{softmax}(\frac{\mathcal{F}_{\text{audio}}\mathcal{F}_{\text{mask}}^T}{\sqrt{D/n_{\text{head}}}})
\end{equation}
\begin{equation}
    \hat{\mathcal{F}}_{\text{mask}} = \mathcal{F}_{\text{mask}} + \mathcal{F}_{\text{mask}} \odot \omega
\end{equation}
where $n_{\text{head}}=8$ and $\odot$ represents element-wise multiplication.

\noindent\textbf{Transformer Decoder:} The decoder utilizes the mixed query $\mathcal{Q}_{\text{mixed}}$ as input and processes multi-scale visual features as key/value pairs. Through the decoding process, output queries $\mathcal{Q}_{\text{output}}$ continuously aggregate visual features and combine with audio information. The final mask is generated through:
\begin{equation}
    \mathcal{M} = \text{FC}(\hat{\mathcal{F}}_{\text{mask}} + \text{MLP}(\hat{\mathcal{F}}_{\text{mask}} \cdot \mathcal{Q}_{\text{output}}))
\end{equation}
where MLP and FC layers integrate different channels to produce the final segmentation prediction.

\noindent\textbf{Loss Functions:} The original AVSegFormer~\cite{gao2024avsegformer} employs two complementary loss terms for training:
\begin{itemize}
\item $\mathcal{L}_{\text{IoU}}$ is a Dice loss comparing predicted segmentation masks with ground truth masks. This loss is particularly effective for handling the class imbalance inherent in AVS tasks where sounding objects often occupy small portions of the frame.
\item $\mathcal{L}_{\text{mix}}$ supervises the audio-visual mixer by computing Dice loss between a predicted binary mask (aggregated from mixed features) and combined foreground labels. This loss enhances the model's ability to handle complex scenes with multiple sound sources.
\end{itemize}

\subsubsection*{D.2 Learning with Balanced Audio-Visual Pairs, Classifier-Guided Feature Alignment}

\begin{figure}[t]
    \centering
    \includegraphics[width=\linewidth]{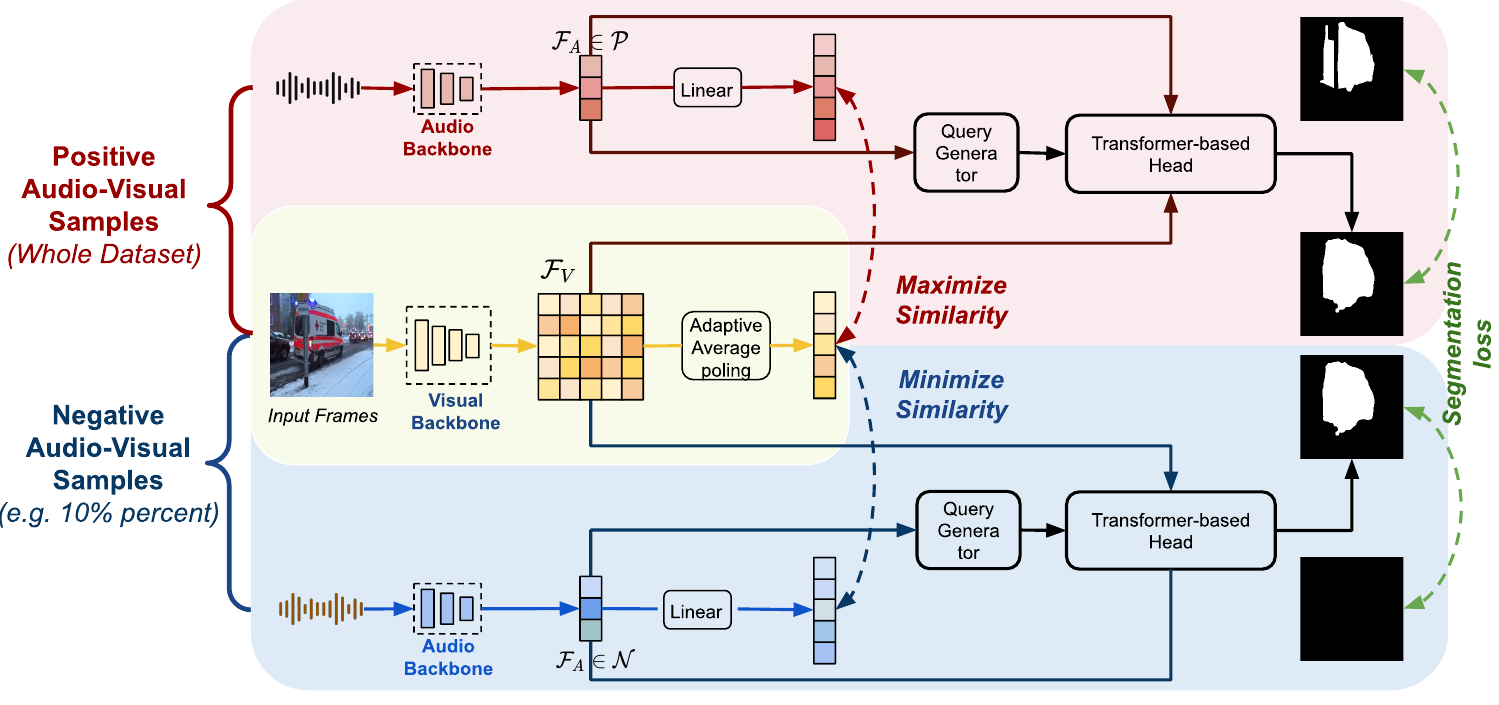}
    \caption{Overview of our robust AVS framework based on AVSegformer~\cite{gao2024avsegformer}. The model processes both positive and negative audio-visual pairs through separate encoders, employs classifier-guided similarity learning for feature alignment, and integrates the features for mask prediction. Positive pairs maximize similarity while negative pairs minimize it, enabling effective discrimination between valid and invalid audio-visual correspondences.}
    \label{fig:framework}
\end{figure}

As illustrated in Figure~\ref{fig:framework}, our approach enhances existing AVS architectures with three key components to address the visual bias problem. Given video frames and audio inputs, we first construct balanced positive and negative audio-visual pairs, where positive pairs represent aligned sound sources while negative pairs correspond to scenarios like silence or off-screen sounds. The model processes these pairs through separate visual and audio encoders to extract modality-specific features. These features undergo similarity-based alignment, optimized through classifier guidance in a contrastive manner. Finally, the aligned features are through the Query Generator module and the Transformer-based Head part before generating the final segmentation masks.

The framework is designed to maintain high segmentation accuracy for positive pairs while effectively suppressing predictions for negative pairs. Positive pairs are trained to maximize feature similarity, while negative pairs minimize it. By processing both types of pairs through this pipeline, the model learns to distinguish between valid and invalid audio-visual correspondences, enabling more robust segmentation in complex real-world scenarios.

Following Section~\ref{sec:method} of the main paper, we implement balanced training by maintaining a 10\% ratio of negative pairs during training. For classifier-guided similarity learning, we compute cosine similarity between aligned audio and visual features:
\begin{equation}
    s(\mathcal{F}_A, \mathcal{F}_V) = \text{cos}(\hat{\mathcal{F}}_A, \hat{\mathcal{F}}_V)
\end{equation}

The binary cross-entropy loss guides similarity learning:
\begin{equation}
\begin{split}
   \mathcal{L}_{\text{BCE}} = -\frac{1}{|\mathcal{P}| + |\mathcal{N}|} & \sum_{j=1}^{|\mathcal{P}| + |\mathcal{N}|}  \left( y_j \log \sigma(s_j) \right. \\
   & \left. + (1 - y_j) \log (1 - \sigma(s_j)) \right),
\end{split}
\end{equation}

The final training objective combines three complementary components:
\begin{equation}
\mathcal{L}_{\text{total}} = \mathcal{L}_{\text{IoU}} + \lambda_1\mathcal{L}_{\text{mix}} + \lambda_{\text{BCE}}\mathcal{L}_{\text{BCE}}
\end{equation}
where $\mathcal{L}_{\text{BCE}}$ guides the similarity learning between audio and visual features, helping the model distinguish between valid and invalid audio-visual correspondences through contrastive learning. The weighting factors $\lambda_1=0.1$ and $\lambda_{\text{BCE}}=1.0$ balance the contributions of each loss component.

\subsubsection*{D.3 Training Details}
Following AVSegFormer~\cite{gao2024avsegformer}'s setting, we adjust the input resolution to $512 \times 512$ to better capture detailed visual information. This model's training employs the AdamW optimizer~\cite{loshchilov2017adamw}, with a batch size of 1 and an initial learning rate of $2 \times 10^{-5}$. The encoder and decoder consist of 6 layers each, with an embedding size of 256. The training protocol extends to 60 epochs for the MS3 dataset and 30 epochs for the S4 dataset, conducted on an NVIDIA RTX A6000 GPU.

\end{document}